%% file: custom.tex
\pdfoutput=1

\documentclass[11pt]{article}

\usepackage[final]{acl}
\usepackage{placeins} 
\usepackage{times}
\usepackage{latexsym}

\usepackage[T1]{fontenc}

\usepackage[utf8]{inputenc}

\usepackage{microtype}

\usepackage{inconsolata}

\usepackage{graphicx}

\usepackage{microtype}
\usepackage{graphicx}
\usepackage{subfigure}
\usepackage{booktabs} 

\usepackage{hyperref}       %
\hypersetup{colorlinks,linkcolor={red}} \usepackage{xspace}

\usepackage{amsmath}
\usepackage{amssymb}
\usepackage{mathtools}
\usepackage{amsthm}
\usepackage{enumitem}
\usepackage{pifont}
\usepackage{hyperref}
\usepackage{url}
\usepackage[most]{tcolorbox}
\usepackage{wrapfig}
\usepackage{float}
\usepackage{threeparttable}
\usepackage{colortbl}
\usepackage{color}
\usepackage{multirow}
\usepackage{graphicx}
\usepackage{arydshln}
\usepackage{caption}
\usepackage{makecell}
\usepackage{subfigure}
\usepackage{bbding}
\usepackage{tabularx}
\usepackage{mathrsfs}
\usepackage{bbm}

\usepackage{listings}

\usepackage[capitalize, noabbrev]{cleveref}

\theoremstyle{plain}

\theoremstyle{definition}

\theoremstyle{remark}

\newcommand{\ourmethod}{{\fontfamily{lmtt}\selectfont \textbf{AgentSymbiotic}}\xspace}
\usepackage{algorithm}
\usepackage{algorithmic}

\usepackage{fix-cm}

\definecolor{customGreen}{rgb}{0.32, 0.64, 0.16}

\usepackage[textsize=tiny]{todonotes}
\title{Symbiotic Cooperation for Web Agents: Harnessing Complementary Strengths of Large and Small LLMs}

\author{Ruichen Zhang\thanks{\ \ Equal contribution.}$^{1}$, Mufan Qiu$^{*1}$, Zhen Tan$^{*2}$, Mohan Zhang$^{1}$, Vincent Lu$^{3}$, Jie Peng$^{1}$,\\ \textbf{Kaidi Xu}$^{4}$\textbf{,} \textbf{Leandro Z. Agudelo}$^{5}$\textbf{,} \textbf{Peter Qian}$^{3}$\textbf{,}
\textbf{Tianlong Chen}$^{1}$ \\
$^{1}$University of North Carolina at Chapel Hill, $^{2}$Arizona State University, \\ $^{3}$Ventus AI, $^{4}$Drexel University, $^{5}$Daice Labs\\
}

\begin{document}
\maketitle
\begin{abstract}
Web browsing agents powered by large language models (LLMs) have shown tremendous potential in automating complex web-based tasks. 
    Existing approaches typically rely on large LLMs (\textit{e.g.}, \texttt{GPT-4o}) to explore web environments and generate trajectory data, which is then used either for demonstration retrieval (for large LLMs) or to distill small LLMs (\textit{e.g.}, \texttt{Llama3}) in a process that remains \textit{decoupled} from the exploration.
    In this paper, we propose \ourmethod, an iterative framework that \textit{couples} data synthesis with task-performance, yielding a \underline{\textit{\textbf{“symbiotic improvement”}}} for both large and small LLMs. Our study uncovers a \textit{complementary dynamic} between LLM types: while large LLMs excel at generating high-quality trajectories for distillation, the distilled small LLMs—owing to their distinct reasoning capabilities—often choose actions that diverge from those of their larger counterparts. This divergence drives the exploration of novel trajectories, thereby enriching the synthesized data. However, we also observe that the performance of small LLMs becomes a bottleneck in this iterative enhancement process. To address this, we propose two \textit{innovations} in LLM distillation: a \underline{\textit{\textbf{speculative data synthesis}}} strategy that mitigates off-policy bias, and a \underline{\textit{\textbf{multi-task learning}}} approach designed to boost the reasoning capabilities of the student LLM. Furthermore, we introduce a \underline{\textit{\textbf{Hybrid Mode for Privacy Preservation}}} to address user privacy concerns. Evaluated on the \textsc{WebArena} benchmark, \ourmethod achieves SOTA performance with both LLM types. Our best Large LLM agent reaches $52\%$, surpassing the previous best of $45\%$, while our 8B distilled model demonstrates a competitive $49\%$, exceeding the prior best of $28\%$. Code will be released upon acceptance.
\end{abstract}

\section{Introduction}
\label{sec:intro}
The autonomous navigation and completion of tasks on the web is a critical capability for AI~\cite{xie2023openagents,yao2023webshop,zhou2023language}. Recent advances in large language models (LLMs) have enabled impressive progress in web browsing agents, as demonstrated by benchmarks such as \textsc{WebArena}~\cite{zhou2023webarena}. Traditionally, current approaches~\cite{su2025learn} adopt a \textit{decoupled} paradigm: First, a data synthesis phase deploys a large LLM to interact with the web environment and generate trajectory data; Subsequently, a task-performing phase uses this data—either as demonstration retrieval for large LLMs or as distillation material for small LLMs.



\begin{figure}[t] 
    \centering
    \includegraphics[width=0.50\textwidth]{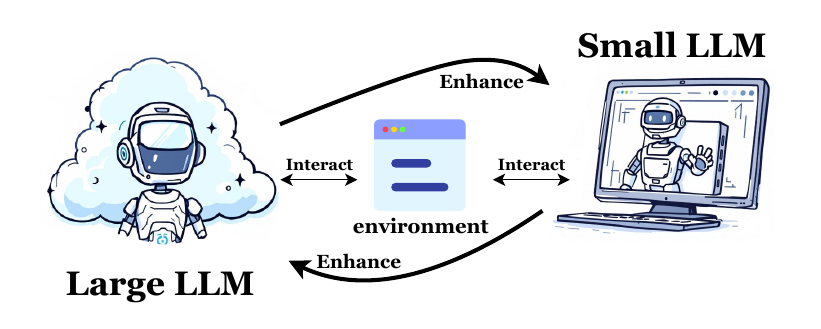} 
    \vspace{-6mm}
    \caption{\small Illustration of the symbiotic improvement between small and large LLMs, where each of them benefits the other.}
    \label{fig:0} 
    \vspace{-3mm}
\end{figure}

In this work, we show that large and small LLMs can engage in a \underline{\textbf{\textit{symbiotic}}} relationship, which enhances both data synthesis and distillation in a \textit{coupled} iterative manner, as illustrated in Figure~\ref{fig:0}. Specifically, we introduce \ourmethod, a novel framework in which large and small LLMs collaborate through an iterative improvement cycle. The process is described as follows:



\begin{figure*}[t] 
    \centering
    \includegraphics[width=\textwidth]{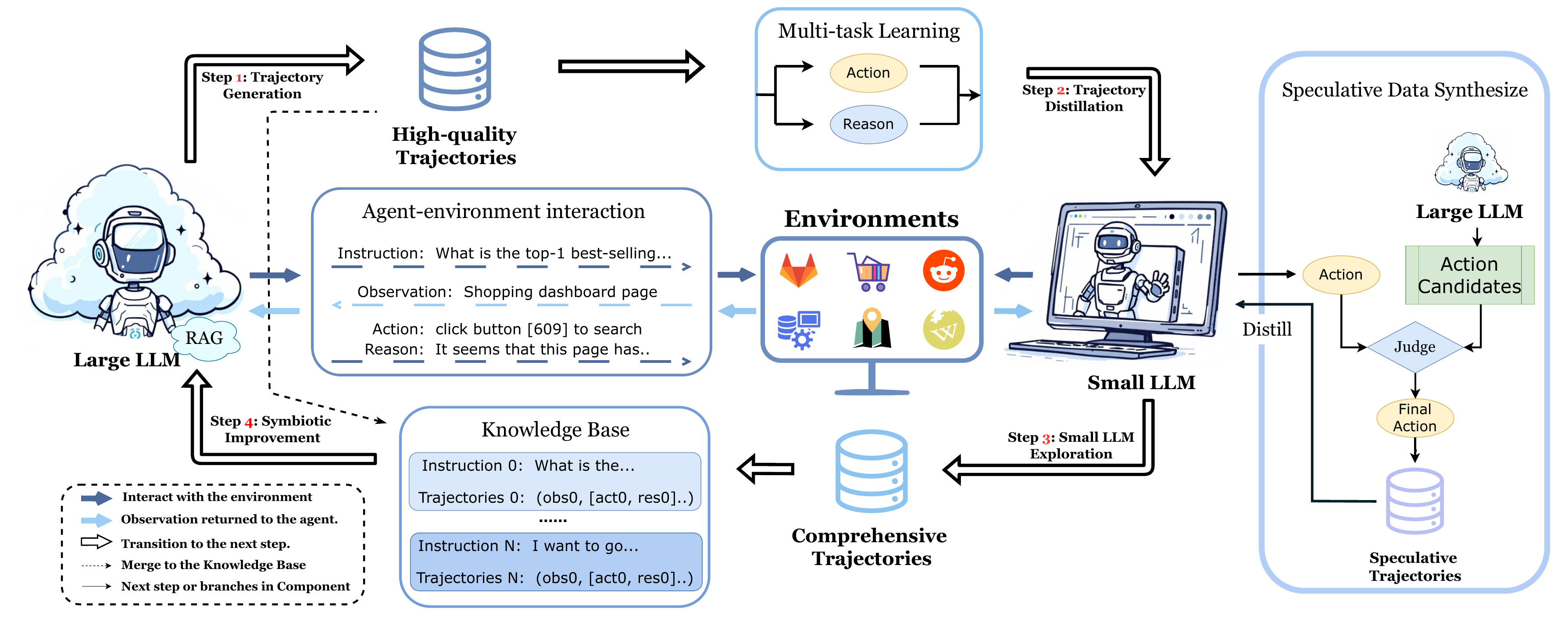}
    \vspace{-5mm}
    \caption{\small Overview of the \ourmethod framework. Step \textcolor{red}{1}: The large LLM interacts with the environment to generate high-quality trajectories, which are then used to distill small LLMs. Step \textcolor{red}{2}: Multi-task learning and Speculative Data Synthesis are applied during distillation to enhance the reasoning capabilities of the small LLM  and mitigate off-policy bias between the two LLMs. Step \textcolor{red}{3}: The small LLM further explores the environment to produce diverse and valuable trajectories. Step \textcolor{red}{4}: Then the knowledge base containing high-quality trajectories and comprehensive trajectories is incorporated into the large LLM's RAG process, improving its performance. This iterative process establishes a symbiotic improvement cycle, enhancing both large and small LLMs over time.}
    \label{fig:1} 
    \vspace{-3mm}
\end{figure*}

    \noindent$\rhd$~\textit{Step} $1$ - \textbf{Trajectory Generation:}
    The large LLM utilizes retrieval-augmented generation (RAG)~\cite{wu2024longmemeval} to refine its performance. By learning from both successful and failed trajectories during rounds of self-interaction, it produces increasingly robust navigation paths.
    
    \noindent$\rhd$~\textit{Step} $2$ - \textbf{Trajectory Distillation:}
    A multi-LLM debate mechanism~\cite{du2023improving,liang-etal-2024-encouraging} is employed to evaluate the generated trajectories. Selected trajectories serve as critical data for distilling small LLMs.
    
    \noindent$\rhd$~\textit{Step} $3$ - \textbf{Small LLM Exploration:}
    Small LLMs, distilled from the large LLM, are deployed to explore the environment more efficiently and extensively due to their faster inference speeds and increased stochasticity in action generation~\cite{sanh2019distilbert}. This process uncovers diverse trajectories—including edge cases or novel solutions that the large LLM might overlook.
    
    \noindent$\rhd$~\textit{Step} $4$ - \textbf{Symbiotic Improvement:} 
     This iterative cycle creates a mutually beneficial loop: the large LLM refines its generation capabilities with enriched feedback from small LLM explorations, while the small LLM continually benefits from high-quality data and distilled expertise provided by the large LLM.

Despite these advances, a significant limitation remains: the performance of small LLMs distilled from large models often falls short of the level required to fully support and enhance large LLMs. Our analysis identifies two root causes behind this gap: (a) \textit{Off-policy bias}~\cite{caccia2024fine} arises when the training data—generated under a large LLM's policy—diverges from the small LLM’s deployment environment; and (b) the loss of critical reasoning capabilities during distillation~\cite{guo2025deepseek} further undermines the small LLM's effectiveness. Based on these insights, we introduce here two key technical innovations in distilling web browsing agents: (a) A \textbf{speculative data synthesis strategy} that mitigates off-policy bias by leveraging multiple action candidates generated by the large LLM to filter and refine the distillation trajectories; and (b) A \textbf{multi-task learning} approach that jointly predicts actions and intermediate reasoning steps, thereby preserving the critical reasoning abilities of the distilled model.


Moreover, as real-world deployments of web agents must safeguard user privacy—particularly when handling sensitive data such as passwords, credit card details, or phone numbers—we integrate a \textbf{hybrid mode for privacy preservation}. In this mode, any step that might involve private data is delegated to a local small LLM rather than a cloud-based large LLM, ensuring confidentiality.
Our contributions can be summarized as follows:
\vspace{-3mm}
\begin{itemize}[leftmargin=*]
\item[\ding{182}] \textbf{\textit{Synergistic Framework.}} We present a novel framework that establishes an iterative, symbiotic cycle between large and small LLMs, enabling them to leverage their complementary strengths for mutual enhancement.
\vspace{-3mm}
\item[\ding{183}] \textbf{\textit{Technical Innovations.}} We introduce two key advancements in distillation techniques: (a) a speculative data synthesis strategy to counteract off-policy bias and (b) a multi-task learning approach to maintain reasoning capabilities.
\vspace{-3mm}
\item[\ding{184}] \textbf{\textit{State-of-the-art Performance.}} Experiments show that on the \textsc{WebArena} benchmark, \ourmethod achieves state-of-the-art performance with both LLM types: the large LLM achieves $52\%$, surpassing the previous best open-source $45\%$, while our 8B distilled LLaMA-3 model achieves $49\%$, approaching the performance of agents based on Claude-3.5.
\vspace{-3mm}

\item[\ding{185}] \textbf{\textit{Hybrid Mode for Privacy Preservation.}} We integrate a hybrid mode for privacy preservation that directs sensitive tasks to a local small LLM, ensuring that private user data remains secure.
\end{itemize}

    \section{Related work}

    \textbf{Web Agents.} LLM-based web agents have gained significant attention in recent years due to their ability to automate, optimize, and enhance a wide range of web-based tasks, such as information retrieval, decision-making, and interactions within dynamic environments~\cite{zhou2023webarena,
    deng2023mind2web, yao2023webshop, pan2024webcanvas, levy2024stwebagentbench,
    dechezelles2024browsergym}.  Many existing approaches~\cite{koh2024tree,
    putta2024agentq, yu2025exact}
    utilize search-based methods like Monte Carlo Tree Search (MCTS) to obtain
    more online examples from the web environments. 
    Although these methods benefit from increased interactions, their performance does not scale with the number of interactions. In contrast, our framework introduces an iterative symbiotic improvement cycle that continually refines both data synthesis and task performance. More detailed related work is provided in Appendix~\ref{app:related}.

    \noindent\textbf{Knowledge Distillation.} Knowledge distillation is a pivotal technique for transferring the advanced capabilities of a large, powerful model to a smaller, more efficient one~\cite{Gou_2021, xu2024survey}. Recent work on LLM distillation~\cite{hinton2015distilling, anand2023gpt4all, hsieh2023distilling} has shifted the focus from merely replicating the teacher model’s output to capturing its underlying reasoning and decision-making processes. Additionally, several methods have been proposed to fine-tune language models for web tasks~\cite{yin2024agent,hong2024cogagent,lai2024autowebglm}, further enhancing decision-making abilities. Despite these efforts, distilled small LLMs still lag behind their larger counterparts in performance. Our work addresses this gap by introducing a novel distillation approach that leverages iterative symbiotic improvements to enhance the capabilities of small LLMs. More detailed distillation related work is provided in Appendix~\ref{app:related}.

\section{Methodology}
\label{sec:method}

\textbf{Overview}. In this section, we introduce our framework, \ourmethod, which is designed to enhance the capabilities of both large and small LLMs. Our approach has two key components: (1) large LLMs access more comprehensive and diverse references through Speculative Data Synthesis during Retrieval-Augmented Generation (RAG)~\cite{wu2024longmemeval}, and (2) small LLMs integrate reasoning into their predictions during the distillation process using synthesized data. 

\textbf{Problem Formulation.} Let $o \in \mathcal{O}$ represent an observation, which consists of an accessible tree structure provided by the web environment along with a corresponding instruction. An action $a \in \mathcal{A}$ corresponds to a command that can be executed and interpreted by the web environment. The reason $r \in \mathcal{R}$ captures the rationale behind why an LLM chooses to execute a specific action in response to an observation. The state $s \in \mathcal{S}$ corresponds to the current state of the environment. At each step $i$, large LLM ($M_{\mathrm{L}}$) predicts the next action $a_{i}$ and reason $r_{i}$ based on the interaction history $\mathcal{H} = (o_{0}, a_{0}, o_{1}, a_{1}, \dots, o_{i})$. In contrast, the small LLM ($M_{\mathrm{S}}$) makes its predictions for $a_{i}$ and $r_{i}$ based solely on the current observation $o_{i}$.

    \subsection{Large LLMs and Small LLMs Can Benefit Each Other}
    \label{sec:LS}


    
    {\textbf{Exploration-Exploitation Trade-off}}. The fundamental difference between large and small LLMs in web-based tasks can be understood through the lens of the \emph{exploration-exploitation} trade-off.
    Large LLMs excel at \emph{exploitation}—that is, they are very effective at accurately selecting actions in well-understood scenarios—while small LLMs, with their faster inference speeds, are more agile and capable of \emph{exploring} a broader range of possible actions.  
    Formally, for a given task $\mathcal{T}$,
    let $E(M, \mathcal{T})$ denote the expected performance of an LLM $M$ on
    tasks drawn from $\mathcal{T}$. Empirically, large LLMs $M_{L}$ often
    satisfy the inequality:
    \begin{equation}
        E(M_{\mathrm{L}}, \mathcal{T}) \;>\; E(M_{\mathrm{S}}, \mathcal{T}), \label{eq:llm-better-smaller}
    \end{equation}
    where $M_{\mathrm{S}}$ represents a small LLM with significantly fewer parameters. 
    
    Moreover, small LLMs tend to be more flexible in their action selection because they are more sensitive to changes in observations and less capable of capturing complex patterns~\cite{wang2024comprehensive}. This increased sensitivity leads them to exhibit more variable behavior, enabling them to explore a larger subset of state-action pairs in the environment. In other words, if we denote by $\pi_{M_L}$ and $\pi_{M_S}$ the policies of the large and small LLMs respectively, small LLMs tend to cover more of the state-action space $(\mathcal{S} \times \mathcal{A})$,
    under a given time or computational budget. This broader coverage can uncover diverse trajectories—including corner cases, failed attempts, or novel solutions—that might be missed by a $M_{L}$ operating on its own.


\noindent{\textbf{Synergistic Gains.}} Large and small LLMs have a \emph{complementary} relationship that can be harnessed in an \emph{iterative} manner, to achieve performance beyond what either model could reach individually. We define Success Rate (SR) as the proportion of tasks in which an LLM reaches the correct goal state, as follows:
    \begin{equation}\label{eq:sr}
        \text{SR}(M) \;=\; \frac{\sum_{i=1}^{N} \mathbb{I}\Bigl(s_{T}^{i} = s_{T}^{\text{goal}}\Bigr)}{N}
        ,
    \end{equation}
    More precisely, $N$ is the total number of tasks, $s_{T}^{i}$ is the final state
    reached by the LLM $M$ after executing the sequence of actions
    $(a_{1}, a_{2}, \dots, a_{T})$ for task $i$, and $s_{T}^{\text{goal}}$ is the
    correct goal state for the task. The indicator function $\mathbb{I}(\cdot)$
    evaluates to $1$ if the final state $s_{T}^{i}$ matches the goal state
    $s_{T}^{\text{goal}}$, and $0$ otherwise, for a time horizon~$T$.
    Consider a scheme with the following components:
    \vspace{-2mm}
    \begin{enumerate}[leftmargin=*,itemsep=1.5pt]
        \item [\ding{182}] The large LLM $(M_{\mathrm{L}})$ interacts with the environment to produce \emph{high-quality
            trajectories}, which are then used to distill the small LLM $(M_{\mathrm{S}})$.
\vspace{-2mm}
        \item [\ding{183}] The distilled $M_{\mathrm{S}}$ subsequently engages in \emph{exploratory interactions}, discovering \emph{new trajectories} that the $M_{\mathrm{L}}$ may have overlooked.
\vspace{-1mm}
        \item [\ding{184}]These additional trajectories are incorporated into the knowledge base
            for the large LLM’s RAG, thereby improving $M_\mathrm{L}$ for the \emph{next iteration} of environment interaction.
    \end{enumerate}
    This process is repeated over multiple rounds, creating a compounding feedback loop where each LLM benefits from the strengths of the other. Let
    $\text{SR}^{\text{(iter)}}(M_{\mathrm{L}}, M_{\mathrm{S}})$ denote the final success rate of
    this \emph{iterative} procedure after several rounds. We then define a \emph{synergy
    metric}~$\Delta$:
    \begin{equation}\small
        \Delta \;=\; \text{SR}^{\text{(iter)}}(M_{\mathrm{L}}, M_{\mathrm{S}}) \;-\; \max\Bigl(\text{SR}
        (M_{\mathrm{L}}),\,\text{SR}(M_{\mathrm{S}})\Bigr) \label{eq:synergy_iterative}
    \end{equation}
    A strictly positive $\Delta$ indicates that the iterative scheme yields higher success rates than the best single-LLM approach, demonstrating the power of synergistic cooperation between both LLM types.

    \vspace{-1mm}
    \subsection{Build RAG-Enhanced Large LLM}
    \label{sec:rag} While some agents enable LLMs to act in augmented observation-action spaces or use search algorithms for web navigation, these methods often face limitations—such as time-consuming design of action spaces, increased interaction steps, and the inability to improve performance iteratively. To overcome these challenges, we enhance the performance of large LLMs by integrating RAG within the \textsc{AgentOccam} framework. The detailed implementation steps are outlined in Algorithm~\ref{alg:rag_large_model}. 


    \noindent{\textbf{Data Synthesis.}}
    Our agent begins by interacting with the environment across various tasks, accumulating both successful and failed trajectories. Each trajectory (denoted as $\mathcal{H}$)
    is decomposed into all possible subsequences that start and end with an observation. For example, a trajectory like
    $(o_{0}, a_{0}, o_{1}, a_{1}, o_{2})$ is split into subsets such as
    $(o_{0}, a_{0}, o_{1})$ and $( o_{1}, a_{1}, o_{2})$, among others. Inspired by recent approaches that use LLMs as judges~\cite{li2024generation,gu2024survey,tan2024large}, we employ a \textit{multi-LLM debate}~\cite{liang-etal-2024-encouraging} mechanism to generate task instructions and summaries for these trajectories. The generated content is evaluated against predefined criteria; only the trajectories that satisfy all criteria are retained. These validated trajectories, along with their corresponding instructions and summaries, are stored in a knowledge base for later retrieval.
\begin{figure*}[htbp] 
    \centering
    \includegraphics[width=\textwidth]{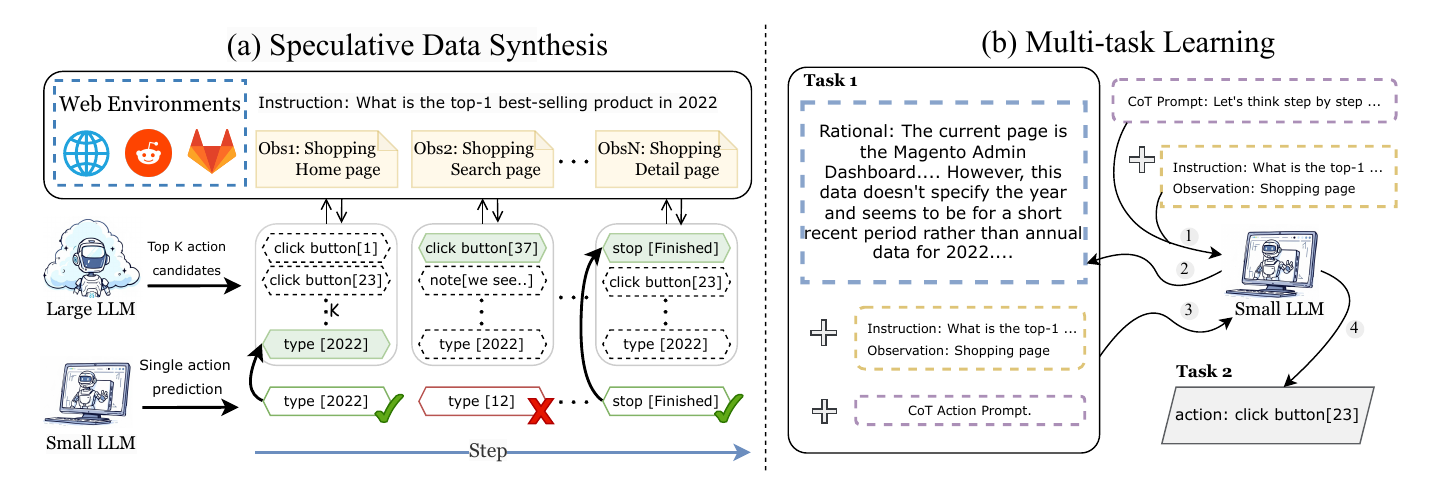}
    \vspace{-10mm}
    \caption{Overview of two key innovations in LLM distillation: (a) \textbf{Speculative Data Synthesis}, which mitigates off-policy bias by leveraging both large and small LLMs. At each step, the small LLM generates an action based on the observation, while the large LLM produces a set of top-$K$ action candidates. If the small LLM's action is within the large LLM's top-$K$ actions, it is accepted (\textcolor{customGreen}{$\checkmark$}); otherwise, the large LLM's action is chosen for subsequent interactions (\textcolor[HTML]{E51400}{\ding{55}}). (b) \textbf{Multi-task Learning}, which enhances reasoning capabilities by training small LLM to predict both actions and rationales, enabling it to handle multiple tasks and address missing reasoning capabilities during distillation. CoT indicates Chain-of-Thought~\cite{wang2024chain}.}
    \label{fig:2} 
    \vspace{-5mm}
\end{figure*}

    \noindent{\textbf{RAG Example Retrieval Strategies.}}
To retrieve relevant knowledge for the agent, we propose a mixture of three strategies:
(a)~\textit{Task-Guided Summary Retrieval}: Queries are generated from task instructions and webpage observations to retrieve relevant past experiences from the RAG knowledge base.
    (b)~\textit{Direct Observation and Instruction Matching}: The current observation and instruction are directly matched with entries in the knowledge base.
    (c)~\textit{Trajectory Similarity Search}: Similar interaction examples are retrieved by computing and comparing trajectory embeddings using cosine similarity. After retrieving $K$ trajectory examples, a {filtering step} performed by an LLM ensures their quality and relevance. The detailed prompt is provided in Appendix~\ref{app:Filter}. Since similarity alone doesn't guarantee usefulness for action prediction, an LLM employs a {chain-of-thought}~\cite{wang2024chain} reasoning process to evaluate and rank these examples. The selected high-quality examples then aid the large LLM in its decision-making during interaction with the environment.

    \vspace{-1mm}\subsection{Improved Distillation for Small LLMs}
    \label{sec:distill} To further enhance the performance of distillation, we introduce two key innovations for processing web-browsing trajectories (see Figure~\ref{fig:2}):
    (a) a speculative data synthesis strategy designed to correct off-policy
    bias~\cite{caccia2024fine}, and (b) a multi-task learning approach aimed at improving reasoning capabilities of the small LLM. These innovations are detailed in Algorithm~\ref{alg:improved_distillation}
    
    \noindent{\textbf{Speculative Data Synthesis.}} 
    In complex environments such as the web, where decision-making tasks require nuanced reasoning and adaptability, it has been observed that a distilled small agent (e.g., a LLaMA-based model) often produces actions that differ significantly from those of a large LLM. Traditional knowledge distillation (KD) methods, such as supervised KD, struggle in such settings due to the large gap in reasoning and decision-making capabilities between the small and large LLMs. To address these challenges, we propose a dynamic collaboration between the large (teacher) and small (student) LLMs to iteratively and adaptively generate high-quality training data.

    Specifically, at each step $i$, the speculative synthesis process unfolds as: \ding{182} \textit{Student Proposal:} The small LLM generates a reasoning trace and an action based on the current observation $o_{i}$ and instruction $I$. \ding{183} \textit{Teacher Evaluation:} Simultaneously, the large LLM, guided by chain-of-thought prompting, generates $K$ action candidates along with their corresponding reasoning traces. \ding{184} \textit{Action Candidate Filtering:} The action proposed by the small LLM is evaluated against the set of candidate actions produced by the large LLM. If the student’s action is among the teacher’s candidates, it is deemed reliable and is executed; if not, it is rejected and replaced by the most reliable candidate selected from the teacher’s proposals.

    This speculative interaction ensures that the small LLM learns to align its policy with the large LLM’s superior reasoning while retaining the ability to explore and propose its own actions. Importantly, the speculative synthesis process evolves over time as the small LLM improves:
    \vspace{-3mm}
    \begin{itemize}[leftmargin=*]
        \item \textbf{Early Training Phase:} During the initial stages, the small LLM often proposes suboptimal or low-quality actions. In this phase, the \textit{Action Candidate Filtering} process closely resembles supervised KD, where the teacher LLM corrects the student by replacing their actions with high-confidence actions from the teacher.
\vspace{-2mm}
        \item \textbf{Later Training Phase:} As the small LLM’s reasoning and action selection performance improves, the filtering process transitions to a more \textit{speculative mode}, where the teacher increasingly accepts the small LLM’s proposals. 
    \end{itemize}
\vspace{-2mm}
    By gradually shifting responsibility from the teacher LLM to the student LLM, this approach enables the small LLM to effectively bridge its knowledge gap with the large LLM.

    The trajectories generated through the speculative data synthesis process are combined with those produced by the large LLM. This combined dataset forms the foundation for further distillation, enabling the small LLM to learn from both its own exploratory actions and the large LLM’s high-quality trajectories. This iterative refinement process results in a small LLM that not only mimics the large LLM’s performance but also demonstrates robust decision-making in diverse environments.

    \noindent{\textbf{Multi-Task Learning.}} 
    Training an agent for complex environments often requires large amounts of manually annotated data, making it challenging to effectively associate observations with the corresponding actions. To overcome this, we leverage a large LLM to generate both action predictions and intermediate reasoning steps, which serve as rich supervisory signals for training small LLMs. In this way, the small LLM learns not only to predict actions but also to produce intermediate reasoning steps that explain those actions.
    
    Specifically, the large LLM is prompted to produce outputs that include \ding{182} \textit{Action Generation:} It predicts the next action based on the agent’s current state. \ding{183} \textit{Full Reasoning:} It generates the rationale explaining why the action is appropriate.


\vspace{-3mm}
\subsection{Hybrid Mode for Privacy Preservation}
\label{sec:privacy}
\vspace{-1mm}

While the large LLM offers superior reasoning and broader knowledge, many web tasks involve confidential or high-stakes information (e.g., passwords and payment details) that must be handled with care. To safeguard user privacy, we propose a hybrid mode that allows the system to automatically switch between a local small LLM and a cloud-based large LLM. 

This hybrid mode operates as follows:

\noindent\textbf{Privacy Detection.}
Before processing any environment observation or action, the content is scanned by a local \texttt{DeepSeek-R1}~\cite{guo2025deepseek} model, which flags potential private information—such as personally identifiable data or security tokens.

\noindent\textbf{Local Processing.}
If the observation or action is deemed private, the decision-making step is delegated to the small LLM deployed locally.

\begin{figure}[t] 
    \centering
    \includegraphics[width=0.48\textwidth]{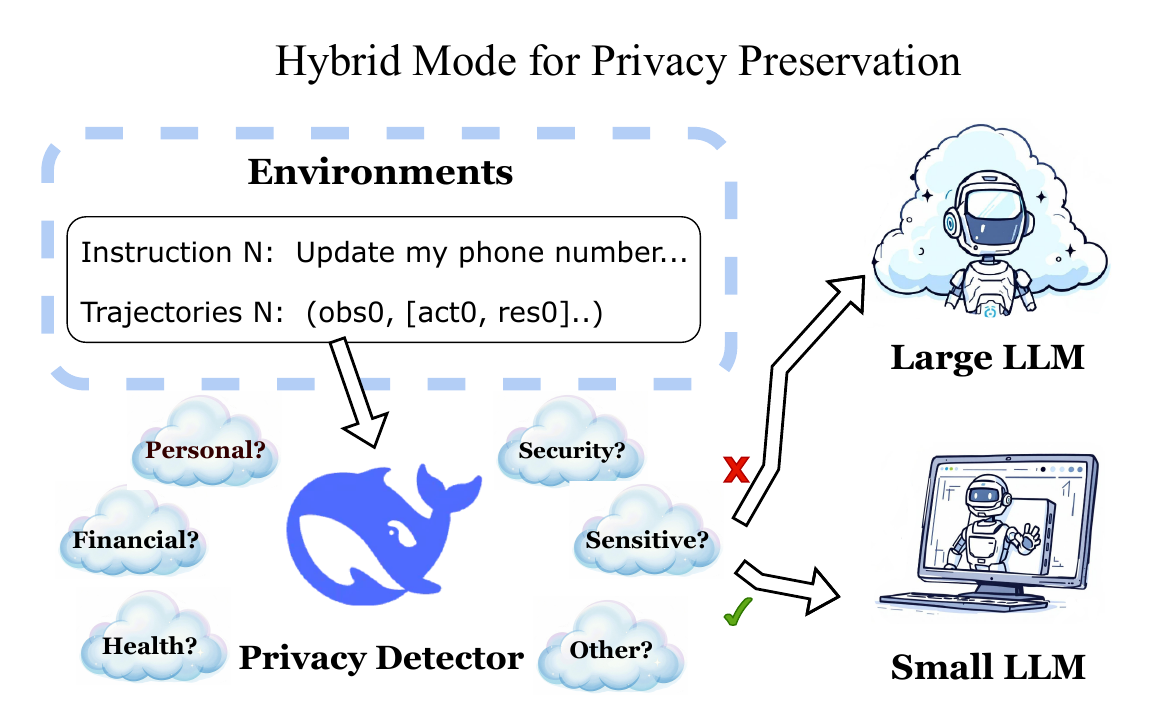} 
    \vspace{-3mm}
    \caption{The Privacy Detector analyzes each step's observation and action for private data. If detected, a local small LLM ensures confidentiality by predicting the next action and reason. Otherwise, a cloud-based large LLM handles predictions, leveraging its superior reasoning capabilities for non-sensitive tasks.}
    \label{fig:3} 
    \vspace{-3mm}
\end{figure}

\noindent\textbf{Cloud Processing.}
If no private information is detected, the agent leverages the large, cloud-based LLM to benefit from its advanced capabilities.

By combining on-device inference for sensitive steps with cloud-based reasoning for non-sensitive tasks, this hybrid mode offers a practical and robust solution for building privacy-preserving web-based agents. Detailed prompts for privacy detection are provided in the Appendix~\ref{app:private}.

    
\vspace{-2mm}
    \section{Experimental Setup}
    \label{sec:exp_setup} \vspace{-1mm}
Our experiment settings are as follows. Implementation details are presented in Appendix~\ref{app:exp}.

    \noindent\textbf{Environment.} \textsc{WebArena} is a benchmark simulating realistic websites across various domains such as e-commerce, collaborative software development, and social forums. Each domain poses a distinct set of tasks (\textit{e.g.}, purchasing items, creating an issue on {\fontfamily{pcr}\selectfont GitLab}, participating in a {\fontfamily{pcr}\selectfont Reddit} discussion), thereby testing the agent’s ability to plan and execute complex, multi-step actions. We report the average \emph{success rate} (SR, defined in Eq.\eqref{eq:sr}) across all 812 tasks as our primary metric, consistent with prior work~\cite{zhou2023webarena}.

    \noindent\textbf{Agents.}
    Unless otherwise stated, we consider two classes of LLMs: (\textit{i}) {Large, closed-source LLMs} (\texttt{Claude-3.5-sonnet}, \texttt{GPT-4-Turbo}, \texttt{GPT-4o}); (\textit{ii}) {Smaller, open-source LLMs}, which include two subcategories: \texttt{DeepSeek-R1-Distill-Qwen-32B} used for privacy detection, and  \texttt{DeepSeek-R1-\\Distill-Llama-8B}, \texttt{Llama-3.2-1B-Instruct}, and \texttt{Llama-3.1-8B-Instruct} for distillation. Large LLMs are accessed via API. All the small LLMs are deployed locally.

\noindent\textbf{Baselines.}
We compare our \ourmethod approach against several representative baselines:  {{Vanilla prompting}}: Use predefined action options to interact with the environment and generate structured responses. Existing baselines explore various approaches to enhancing LLM-based web agents, including optimization, adaptation, policy learning, planning, workflow memory, API integration, evaluation, and multi-agent strategies {{AgentOccam}}~\cite{yang2024agentoccam}, {{Learn-by-Interact}}~\cite{su2025learn}, {WebArena-replication}~\cite{zhou2023webarena}, {SteP-replication}~\cite{sodhi2024step}, LATS~\cite{zhou2023language}, AWM~\cite{wang2024agent}, API-Based Agent~\cite{song2024beyond}, AutoEval~\cite{pan2024autonomous}, WebPilot~\cite{zhang2024webpilot}. See more detailed baseline in Appendix~\ref{app:base}.


\begin{table}[t]
\centering
\caption{Comparison of final success rates (SR) among various large  LLM  and small LLM base agents on \textsc{WebArena}. Scores marked with $^{*}$ indicate cited scores from the corresponding papers' experiment scores.}
\label{tab:main_results}
\vspace{-3mm}

    \resizebox{0.48\textwidth}{!}{
    \begin{tabular}{l c c}
    \toprule
    \textbf{Method} & \textbf{Model} & \textbf{SR (\%) $\uparrow$} \\ 
    \midrule
    WebArena-replication  & \texttt{GPT-4-Turbo}         & $16.5^{*}$ \\     {AutoEval}             & \texttt{GPT-4}     & $20.2^{*}$ \\
    {Reflection}             & \texttt{Claude-3.5}     & $32.4^{*}$ \\
    SteP-replication      & \texttt{GPT-4-Turbo}         & $33.3^{*}$ \\
    {LATS}             & \texttt{Claude-3.5}     & $34.2^{*}$ \\
    {AWM}             & \texttt{GPT-4}     & $35.6^{*}$ \\
    {WebPilot}             & \texttt{GPT-4o}     & $37.2^{*}$ \\
    {Learn-by-Interact}             & \texttt{Claude-3.5}     & $39.2^{*}$ \\
    {API-Based Agent}             & \texttt{GPT-4o}     & $43.9^{*}$ \\
    {AgentOccam}            & \texttt{GPT-4-Turbo}    & $45.7^{*}$ \\
    {AgentOccam}             & \texttt{Claude-3.5}     & $48.5$ \\
    \rowcolor[gray]{0.90}\textbf{\ourmethod}     & \texttt{Claude-3.5}     & $\textbf{52.1}$ \\

    \midrule
    Vanilla prompting & \texttt{LLaMA-1B}       &  $2.4$ \\
    Vanilla prompting & \texttt{LLaMA-8B}       &  $5.6$ \\
    Vanilla prompting & \texttt{DeepSeek-R1-8B}       &  $8.5$ \\
     {Learn-by-Interact}     & \texttt{Codegemma-7B}            & $17.9^{*}$ \\
    {Learn-by-Interact}     & \texttt{Codestral-22B}            & $28.0^{*}$ \\
    \rowcolor[gray]{0.90}\textbf{\ourmethod}      & \texttt{LLaMA-1B}       & $24.1$ \\
    \rowcolor[gray]{0.90}\textbf{\ourmethod}      & \texttt{DeepSeek-R1-8B}       & $43.6$ \\
    \rowcolor[gray]{0.90}\textbf{\ourmethod}& \texttt{LLaMA-8B}       & $\textbf{48.5}$ \\

    \midrule
    \textbf{\ourmethod-{\fontfamily{lmtt}\selectfont \textbf{Hybrid}}}         & \makecell{\texttt{Claude-3.5} \\+ \texttt{LLaMA-8B}}     & $50.5$ \\

    \bottomrule
    \end{tabular}
    }
   
\end{table}

\vspace{-2mm}
\section{Experiment Results}
    \label{sec:experiments}
\subsection{Superior Results of \textbf{\ourmethod}}
\label{sec:main_results}
\vspace{-1mm}

Table~\ref{tab:main_results} summarizes the final success rates of each agent on the entire \textsc{WebArena} (\#$812$ tasks). We obtain the following findings:
\ding{182}\textbf{\ourmethod{\fontfamily{lmtt}\selectfont \textbf{-Claude-3.5}}} achieves an success rate (SR) of \textbf{$52\%$}, significantly outperforming the previous best open-source result of (\textbf{$45\%$}). This improvement underscores the effectiveness of our iterative synergy approach (\S\ref{sec:LS}) and the use of diverse, high-quality trajectories in the RAG module (\S\ref{sec:rag}).
\ding{183}\textbf{\ourmethod{\fontfamily{lmtt}\selectfont \textbf{-LLaMA-8B}}} attains \textbf{$48.5\%$}, a substantial improvement over the original \texttt{LLaMA-8B} baseline ($5\%$) and prior small-LLM methods (up to $28\%$). 
    This validates our claim that improved \emph{distillation} (\S\ref{sec:distill}) and re-using newly explored trajectories enable even 8B-parameter models to approach large-LLM performance (\textbf{$52\%$}). 
\vspace{-1mm}
\ding{184} Our \textbf{\ourmethod} consistently maintains a clear performance gap over standard few-shot or fine-tuned approaches. This highlights that the \emph{complementary dynamic} between large and small LLMs—exploited in a multi-round loop—is critical for robust performance on complex web tasks. One of the trajectory examples is shown in Appendix~\ref{app:Trajectory}.

Figure~\ref{fig:6} presents our iterative experiment on a subset of \textsc{WebArena} tasks. It shows that as the iteration number increases, the synergy metric ($\Delta$), as defined in Equation~\ref{eq:synergy_iterative}, also gradually increases, indicating an improvement in our performance.
\begin{figure}[t] 
    \centering
    \includegraphics[width=0.48\textwidth]{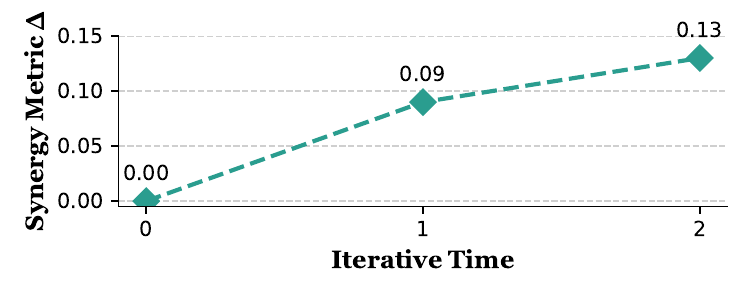} 
    \vspace{-6mm}
    \caption{ The synergy metric ($\Delta$), which defined in Equation~\ref{eq:synergy_iterative}, increases as the iterative time progresses.}
    \label{fig:6} 
    \vspace{-5mm}
\end{figure}

\begin{table*}[t]
\centering
\caption{Ablation study on small LLM distillation for comparison of success rate (SR in $\%$) in specific \textsc{WebArena} sub-domains. (\#Tasks) indicates the number of scenarios in each domain. ``multi-task'' denotes Multi-Task Learning for Reasoning. ``Speculative'' denotes Speculative Data Synthesis.}
\label{tab:ablation}
\vspace{-3mm}

\resizebox{\textwidth}{!}{
\begin{tabular}{l c c c c c c c}
\toprule
\textbf{Agent}  & 
\makecell{\textbf{Overall SR} \\ (\#812)} & 
\makecell{{\fontfamily{pcr}\selectfont Shopping} \\ (\#187)} & 
\makecell{{\fontfamily{pcr}\selectfont {Shopping Admin}} \\ (\#182)} & 
\makecell{{\fontfamily{pcr}\selectfont {GitLab}} \\ (\#180)} & 
\makecell{{\fontfamily{pcr}\selectfont {Map}} \\ (\#109)} & 
\makecell{{\fontfamily{pcr}\selectfont {Reddit}} \\ (\#106)} &
\makecell{{\fontfamily{pcr}\selectfont {Multisite}} \\ (\#48)} \\
\midrule
\texttt{LLaMA-8B}  & $40.8$ & $50.3$ & $30.8$ & $41.1$ & $37.6$ & $51.9$ & $22.9$ \\

\emph{  + multi-task}  & $43.2$ & $46.5$ & $29.1$ & $46.1$ & $45.0$ & $61.3$ & $29.2$ \\

\emph{  + speculative}  & $46.8$ & $46.0$ & $34.6$ & $48.3$ & $56.9$ & $68.9$ & $18.8$ \\

\emph{  + speculative + multi-task}  & $48.5$ & $48.7$ & $41.2$ & $47.2$ & $57.8$ & $63.2$ & $27.1$ \\


\bottomrule

\end{tabular}
}

\end{table*}

\begin{figure*}[t] 
\vspace{-3mm}
    \centering
    \includegraphics[width=\textwidth]{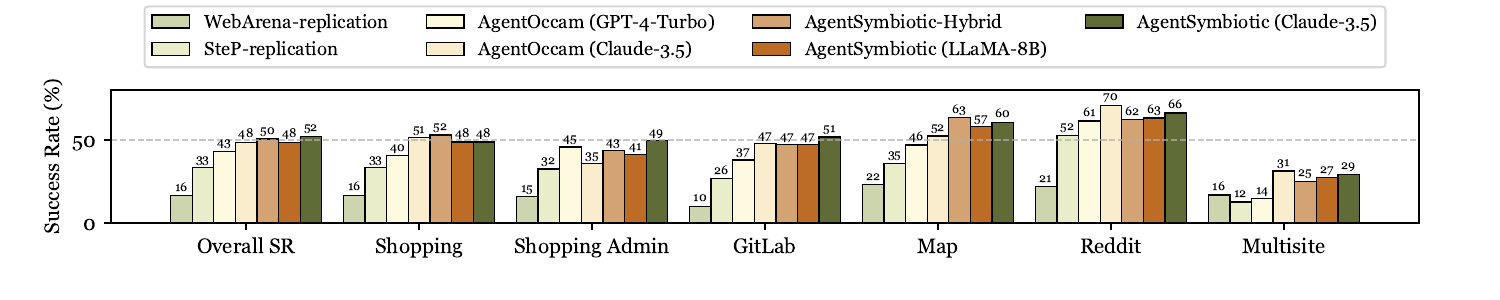}
    \vspace{-10mm}
    \caption{Comparison of success rates between our method and the baseline across different task categories.}
    \label{fig:5} 
    \vspace{-4mm}
\end{figure*}

\subsection{Ablation - Dissecting \textbf{\ourmethod}}
\label{sec:ablation}

We further dissect the gains of our framework by analyzing the two key innovations introduced in the {distillation} process (\S\ref{sec:distill}):
(\textbf{i})~\textit{speculative data synthesis} to mitigate off-policy bias,
(\textbf{ii})~\textit{multi-task learning} to preserve reasoning capabilities.

Table~\ref{tab:ablation} provides the {success rate} of various \texttt{LLaMA-8B} configurations. We compare vanilla supervised fine-tuning (\emph{\texttt{LLaMA-8B}}),  Multi-Task Learning for Reasoning (\emph{multi-task}), and Speculative Data Synthesis (\emph{speculative}) strategy.
Results show that:
\ding{182}~Switching from plain SFT ($40\%$) to “\texttt{LLaMA-8B} \emph{+ speculative}” ($46.8\%$) confers a large boost, validating that “teacher-filtered” data expansions help rectify off-policy mismatches.
\ding{183}~Combining \emph{speculative} with \emph{multi-task} ($49\%$) yields the best performance, reinforcing the importance of maintaining chain-of-thought reasoning while also exploring the environment for speculative data synthesis. 
\ding{184}~We observe that a straightforward fine-tuning approach, without incorporating any \emph{speculative} or \emph{multi-task}, still achieves a remarkably high success rate (SR) of $40\%$, significantly surpassing the $28\%$ SR of the previous $22$B Learn-by-Interact model. This improvement can be attributed to several key factors: (a) our high-quality trajectories generated by a large LLM serve as a distillation dataset, (b) we employ a multi-LLM debate mechanism to select execution trajectories that are valuable for distillation step, and (c) our experiments are built upon the {AgentOccam}  framework, which includes an observation compression component to improve the performance.



\subsection{Domain-Specific Analysis}

To validate that our \textbf{\ourmethod} improvements generalize across different web domains, we report performance per domain in \textsc{WebArena}. 
Figure~\ref{fig:5} shows a representative subset of tasks for {\fontfamily{pcr}\selectfont Shopping, Shopping Admin, GitLab, Map, Reddit}, and  {\fontfamily{pcr}\selectfont Multisite} forums. 
\ourmethod consistently outperforms or closely matches the best domain-specific baselines~(e.g., $7\%$ higher for {\fontfamily{pcr}\selectfont GitLab} tasks), highlighting the advantage of iterative synergy in discovering domain-specific action patterns (especially via small LLM exploration) and systematically incorporating them into the large LLM’s RAG knowledge base and help decision-making.

\noindent\textbf{Qualitative Observations.} 
We observe that tasks in “{\fontfamily{pcr}\selectfont Shopping Admin}”, “{\fontfamily{pcr}\selectfont Shopping}” or “{\fontfamily{pcr}\selectfont GitLab}” often require multi-step forms and error-handling logic. 
In these domains, small LLMs can occasionally stumble onto unorthodox solutions (e.g., toggling unexpected web elements or exploring deeper page links), which subsequently become valuable references in the large LLM’s RAG knowledge store.
Such synergy is precisely the mechanism described in Section~\ref{sec:LS}, wherein small LLM exploration broadens the \emph{action-state coverage}, enabling large LLMs to better \emph{exploit} newly discovered or less conventional paths.

\vspace{-1mm}
\subsection{Hybrid Mode Analysis}
\begin{figure}[t] 
    \centering
    \includegraphics[width=0.48\textwidth]{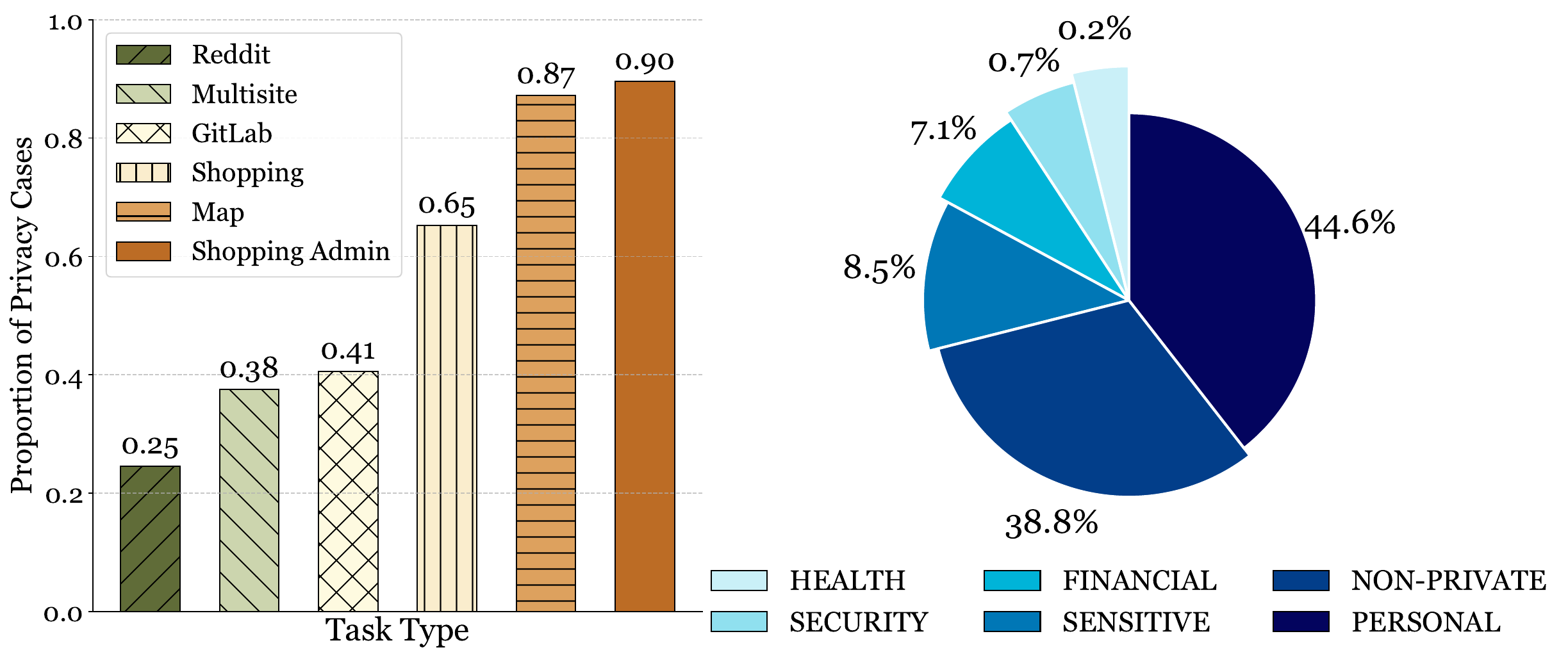} 
    \caption{The Privacy Detector detects and categorizes tasks containing privacy information. We analyze their distribution to understand privacy interactions better.}
    \label{fig:4} 
    \vspace{-5mm}
\end{figure}

We utilize a locally deployed \texttt{DeepSeek-R1} in hybrid mode for privacy preservation to analyze whether each observation and action contains privacy-related information. As shown in Figure ~\ref{fig:4}, we present the probability of encountering privacy-sensitive information across different task types. Experimental results indicate that the {\fontfamily{pcr}\selectfont Shopping Admin} category exhibits the highest occurrence of privacy-related information, primarily due to webpage observations containing sensitive details such as phone numbers, shipping addresses, and purchase histories. In contrast, categories like {\fontfamily{pcr}\selectfont Reddit} and {\fontfamily{pcr}\selectfont GitLab} rarely involve filling in or viewing personal information. 

Furthermore, personal privacy information constitutes a significant portion, reaching $44.6\%$, while the total proportion of privacy-related information sums up to $61.2\%$, far exceeding the $38.8\%$ of non-privacy-related cases. These findings highlight the critical importance of safeguarding privacy information in the domain of autonomous agents.  Privacy detection example is shown in Appendix~\ref{app:hybrid}.

\section{Conclusion}
\label{sec:Conclusion}
In this paper, we introduced \ourmethod, an efficient and straightforward framework that establishes an iterative cycle in which a large LLM and a small LLM continuously enhance each other’s performance. Within this framework, we proposed two novel distillation techniques—speculative data synthesis and multi-task learning—that significantly improve the effectiveness of distilling small LLMs. Additionally, we designed a hybrid mode for privacy preservation, leveraging the complementary strengths of large and small LLMs to safeguard users’ private information.


\clearpage
\section*{Limitations}

\begin{itemize}[leftmargin=*]
    \item \textbf{Budget Constraints:} Due to budget constraints, models like GPT-o1 were not included in our experiments. Moreover, migrating to other benchmarks also incurs substantial API costs. Therefore, similar to many related papers~\cite{yang2024agentoccam,zhou2023webarena,sodhi2024step}, we focus solely on the \textsc{WebArena}~\cite{zhou2023webarena} framework. However, our method is simple and efficient, without any benchmark-specific optimizations, making it easily transferable to other models.
Also, we are unable to measure results after multiple iterative steps and can only evaluate the results after three iterations on a subset of tasks.
Moreover, we were unable to reproduce the results of all baselines. Instead, we referenced the results reported in their respective papers and marked them with an asterisk (*) in the table for clarification.

    \item \textbf{Privacy Assessment:} Since \textsc{WebArena} does not provide labels for private information within its tasks, we are unable to quantitatively assess privacy protection. We did not explore private methods further, evaluate DeepSeek~\cite{guo2025deepseek} on other private datasets, or investigate additional distillation methods, as our primary focus is on the symbiotic improvement of both large and small LLMs.

    \item \textbf{Hardware and Time Constraints:} Extending distillation to more and larger models is highly challenging due to hardware and time limitations. Therefore, we selected these three models \texttt{DeepSeek-R1-Distill-Llama-8B}, \texttt{Llama-3.2-1B-Instruct}, and \texttt{Llama-3.1-\\8B-Instruct} for our distillation experiments.


    \item \textbf{Comparison with Open-Source Methods:} In the \textsc{WebArena} leaderboard results, we compare our method only with projects that have open-source code. The highest previous result among open-source methods was \textit{AgentOccam-Judge}~\cite{yang2024agentoccam}, with a score of 45.7. In this paper, we do not consider closed-source results such as \textit{OpenAI Operator}.

    \item \textbf{Temperature Setting and Variability in Results:} Since LLM outputs often do not strictly follow our instructions to generate structured and multi-faceted responses, we set the temperature to 0.6. As a result, our experimental outcomes may vary across multiple attempts on the same task, sometimes succeeding and sometimes failing. However, due to hardware and budget constraints, we were unable to conduct multiple trials to estimate the variance in accuracy.
\end{itemize}

\section*{Ethics Statement}
Our work on symbiotic web-browsing agents underscores a commitment to responsibly advancing AI while protecting user privacy. Specifically, we use only publicly available benchmark data from \textsc{WebArena} and incorporate a “hybrid mode” that detects and handles sensitive or personal information locally, thus limiting the exposure of private data to external services. Any system built upon our framework should similarly ensure that interactions involving potentially identifiable or private user data are confined to secure, on-device models. Although we demonstrate performance gains through iterative cooperation between large and small LLMs, measures must be taken to guard against possible misuse, such as unauthorized logging of passwords or personal data. Moreover, in adhering to standard research ethics, all code and experiments were designed with transparency in mind, prioritizing reproducibility, explainability, and the principle of “do no harm”.

\section*{Acknowledgement}
This work is generously supported by Ventus AI with Funds and Computing.

\bibliography{custom}

\appendix

\onecolumn
    \section{Related Work}
     \label{app:related}
     \textbf{Agent.} Many agent-based methods have been proposed to tackle real-world challenges, spanning diverse domains such as software engineering, reinforcement learning, multi-agent collaboration, and web interaction.

One line of research focuses on enhancing agent decision-making and problem-solving capabilities. For instance, Monte Carlo Tree Search (MCTS) and Hindsight Feedback have been employed to improve software agents~\cite{antoniades2024swe}, while an MCTS-based approach has been designed to update foundation models for long-horizon tasks~\cite{yu2025improving}. Additionally, bi-level tree search has been explored as a mechanism for self-improving LLM decision-making~\cite{light2025strategist}.

Another key area of advancement is multi-agent collaboration and coordination. Recent work has introduced a novel framework for multi-agent motion generation~\cite{tian2025direct}, as well as an Internet of Agents (IoA) framework that enhances collaboration among autonomous agents using large language models~\cite{chen2025internet}. Similarly, a multi-agent system has been developed to solve complex queries by leveraging specialized agents for different sub-tasks~\cite{li2025agentoriented}, while efficient offline coordination has been explored through diffusion-based trajectory stitching~\cite{yuan2025efficient}.

In the realm of LLM-powered agents, researchers have investigated their application in interactive and visual environments. For example, an embodied agent has been designed to learn causal relationships in the open-world setting of Minecraft~\cite{yu2024adam}, and a new benchmark has been proposed to evaluate coding agents' performance in real-world software engineering tasks involving visual elements~\cite{yang2025swebench}. Additionally, multiple studies have focused on LLM-based web agents, such as leveraging webpage UIs for text-rich visual understanding~\cite{liu2025harnessing} and synthesizing agent trajectories using web tutorials~\cite{xu2025agenttrek}.

Further research has explored LLM alignment and adaptation. Studies have examined the moral alignment of LLM agents~\cite{tennant2025moral}, the differences between aligned LLMs and browser-based agents~\cite{kumar2025aligned}, and strategies for LLM-driven self-improvement in web-based tasks~\cite{patel2025large}. Additionally, a retrieval-augmented generalist agent has been proposed to enable in-context adaptation to new environments~\cite{sridhar2025regent}, while a large-scale benchmark has been introduced to evaluate API-based agents in real-world scenarios~\cite{shen2025shortcutsbench}.

Finally, benchmarking and evaluation frameworks have become increasingly prevalent in agent research. Efforts include benchmarking multimodal retrieval-augmented generation using dynamic VQA datasets~\cite{li2025benchmarking}, assessing long-context multimodal agents via video-based web tasks~\cite{jang2025videowebarena}, and designing an asynchronous planning benchmark for LLM-driven agents~\cite{gonzalez-pumariega2025robotouille}.

As agent research continues to evolve, these developments pave the way for more capable, adaptable, and collaborative intelligent systems across a wide range of real-world applications.

    \textbf{Web Agent.} Recent research has made significant strides in improving web agents, particularly by leveraging curated or automatically synthesized interaction trajectories as training datasets or in-context examples~\cite{qi2024webrl, shen2024scribeagent, hu2024agentgen, zeng2023agenttuning}. For instance, \citet{su2025learn} proposed a data-centric framework that enables LLM agents to adapt to new environments by synthesizing agent-environment interaction trajectories without requiring human annotations.

Another active area of research focuses on multi-agent collaboration for complex web tasks~\cite{hou2024coact, fourney2024magenticone, zhang2024webpilot}. Within this domain, \citet{fourney2024magenticone} and \citet{fu2025agentrefine} introduced a multi-agent architecture where a lead agent is responsible for planning, tracking progress, and dynamically re-planning to recover from errors. Additionally, \citet{yang2024agentoccam} demonstrated that refining a web agent’s observation and action space to better align with LLM capabilities can yield impressive zero-shot performance.

Several studies have incorporated Monte Carlo Tree Search (MCTS) techniques to enhance web agents' decision-making capabilities. These methods iteratively expand intermediate states (tree nodes) through multiple trials on the same task~\cite{zhou2023language, zhang2024webpilot, putta2024agentq}. \citet{koh2024tree} further refined this approach by employing a trained value function to guide the search process and backtrack within the task execution tree. Meanwhile, Auto Eval and Refine~\cite{pan2024autonomous} introduced a reflective reasoning mechanism~\cite{shinn2023reflexion}, using a dedicated evaluator to refine task execution based on insights from previous trials.

Earlier research explored prompt-based methods~\cite{yao2023react, yang2024swe, gur2023real, zhang2023you}, though these approaches are inherently constrained by the capabilities of their underlying foundation models. Other studies have focused on training LLMs using human-annotated examples~\cite{chen2023fireact, li2020mapping}, while recent advancements have introduced progressive understanding web agents for web crawler~\cite{DBLP:journals/corr/abs-2404-12753} and web scraper generation~\cite{DBLP:conf/emnlp/HuangGPLLXWC24}.

Several works have also explored environment modeling and reinforcement learning for web agents. \citet{chae2025web} proposed Web Agents with World Models, which learn and leverage environment dynamics for web navigation. \citet{qi2025webrl} introduced WebRL, a framework that trains LLM web agents using a self-evolving online curriculum reinforcement learning approach. Additionally, \citet{xu2025agenttrek} developed AgentTrek, which synthesizes agent trajectories using web tutorials as guidance.

To benchmark web agent performance, recent studies have introduced TurkingBench~\cite{DBLP:journals/corr/abs-2403-11905}, a challenging benchmark for evaluating web agents across various tasks. Furthermore, research on contextual understanding has led to methods that enhance decision-making by learning to better interpret web pages~\cite{lee2025learning}.

As web agents continue to evolve, these advancements contribute to more adaptive, autonomous, and intelligent systems capable of efficiently navigating and interacting with complex web environments.

    \noindent\textbf{Knowledge Distillation.}  Earlier methods~\cite{hinton2015distilling, kim2020paraphrasing, mirzadeh2019improved, 10.5555/3294771.3294842, fu2023specializing, magister2023teaching, mukherjee2023orca, li2024mixed} focus on training a smaller student network based on the output of a larger teacher network. \citet{huang2022incontext} introduces in-context learning distillation to combine in-context learning objectives with language modeling objectives to distill both the in-context few-shot learning ability and task knowledge to the smaller models.
    Besides, \citet{chen2023fireact} demonstrates that fine-tuning Llama2-7B with 500 agent trajectories generated by GPT-4 leads to a $77\%$ HotpotQA performance increase. \citet{xu2024speculative} shows how to solve off-policy bias. \citet{su2025learn} showcases that utilizing the synthesized trajectory data for training yields better results compared to using it as in-context examples.

\noindent\textbf{Benchmarks.} 
Recent advancements in benchmarking have introduced diverse evaluation datasets targeting specific capabilities of large language models (LLMs). SWE-bench~\cite{jimenez2023swe} focuses on assessing LLMs' performance in software engineering tasks, including code generation, debugging, and documentation. \textsc{WebArena}~\cite{zhou2023webarena} evaluates the ability of LLMs to navigate websites, extract information, and perform web-based tasks in simulated online environments. OSWorld~\cite{xie2024osworld} provides a platform to test LLMs' reasoning and adaptability in open-ended, dynamic, and exploratory simulated worlds. Finally, Spider2-V~\cite{cao2024spider2} extends the original Spider benchmark by introducing more complex SQL queries and diverse database interactions, testing LLMs' proficiency in structured query language tasks. These benchmarks collectively push the boundaries of LLM evaluation across software engineering, web interaction, open-world reasoning, and database management.

\clearpage
\section{Implementation Details} \label{app:exp}
\subsection{Distillation Training}
We conduct the distillation training on \textbf{8 H100 GPUs}, using full-parameter fine-tuning for the models \texttt{Llama-3.2-1B-Instruct}, \texttt{Llama-3.1-8B-Instruct}, and \texttt{DeepSeek-R1-Distill-Llama-8B}. The training process spans \textbf{2 epochs}, with a learning rate of $10^{-4}$ and a context length of $10,000$. The distillation methodology follows the guidelines provided in \texttt{meta-llama/llama-cookbook}. We adopt the \texttt{alpaca\_dataset} format and enable Fully Sharded Data Parallel (FSDP) to facilitate efficient distributed training.

\subsection{Inference Pipeline}
For inference, we employ the \texttt{vLLM} framework, running on \textbf{4 H100 GPUs}. The \textsc{WebArena} framework is deployed on \textbf{8 CPU machines}, utilizing an Amazon Machine Image (AMI) pre-installed with all necessary websites. To enhance efficiency, we leverage the official task-parallel Bash script for parallel execution, rather than processing tasks sequentially by task ID.

\subsection{Chain-of-Thought Prompting}
For Chain-of-Thought (CoT) prompting, we follow the design principles outlined in \textsc{Thinking-Claude},  ensuring structured and effective reasoning in model responses. The implementation details can be found at \href{https://github.com/richards199999/Thinking-Claude}{\texttt{github.com/richards199999/Thinking-Claude}}.

\section{Baselines} \label{app:base}
We compare our \ourmethod approach against several representative baselines:  {{AgentOccam}} ~\cite{yang2024agentoccam}: An LLM-based web agent that refines its observation and action spaces, aligning them more closely with the LLM’s inherent capabilities.
{{Learn-by-Interact}} ~\cite{su2025learn}: A data-centric framework designed to adapt LLM agents to the environment without the need for human annotations.
{WebArena-replication} ~\cite{zhou2023webarena}: An agent that is implemented in a few-shot in-context learning fashion with
powerful large language models.
{SteP-replication} ~\cite{sodhi2024step}: A dynamic framework to compose LLM policies for solving diverse web tasks through adaptable control states.
LATS~\cite{zhou2023language}: A framework integrating reasoning, acting, and planning via Language Agent Tree Search.
AWM~\cite{wang2024agent}: A workflow memory method for guiding agent decision-making.
API-Based Agent~\cite{song2024beyond}: A framework combining API calls and web browsing for web tasks.
AutoEval~\cite{pan2024autonomous}: An evaluation-driven approach for improving web navigation and device control.
WebPilot~\cite{zhang2024webpilot}: A multi-agent system enhancing MCTS for complex web tasks.

    \clearpage
    \section{RAG Algorithm} \label{app:rag}

    \begin{algorithm}
        [htbp]
        \caption{RAG-Enhanced Large LLM}
        \label{alg:rag_large_model}
        \begin{algorithmic}
            [1]

            \STATE \textbf{Input:} Task instruction $I$, current observation
            $o_{i}$, RAG knowledge base $B_{RAG}$, number of retrieved examples
            $K$. \STATE \textbf{Output:} Interaction History$H$.

            \FOR{each instruction $I$ in environment} \STATE $T =interact(I)$
            \STATE $T' = subsequences(T)$ \STATE //Split $T$ into subsequences
            $T'$ \FOR{each $T'$} \STATE $I', S' = LLM(T')$ \STATE //generate task
            instructions and summaries \IF{$LLM(T', I', S')$} \STATE
            $B_{RAG}.append(T', I', S')$ \STATE //Store $(T', I', S')$ in
            $B_{RAG}$ if valid. \ENDIF \ENDFOR \ENDFOR

            $H \gets \emptyset$ \STATE // Initialize History. \FOR{each Step $i$ in interaction for $I$}
            \STATE $q \gets M_{L}(I, o_{i})$ \STATE //Generate retrieval queries
            \STATE $E \gets \emptyset$ \STATE // Initialize retrieved examples. \FOR{strategy in \{retrieval strategies\}}
            \STATE
            $E \gets E \cup \text{Retrieve}(q,{B}_{RAG},\text{strategy}, H_{i-1})$
            \ENDFOR \STATE $E_{\text{filtered}}\gets \text{Filter}(E, M_{L})$

            \STATE $[a_{i}, r_{i}] \gets M_{L}(I, o_{i}, E_{\text{filtered}})$

            \STATE $H_{i+1}\gets H_{i}.append(o_{i},[a_{i}, r_{i}])$ \ENDFOR
            \STATE \textbf{Return:} $H$.
        \end{algorithmic}
    \end{algorithm}
    
Given the WebArena environment, we can first leverage commonly accessible resources such as official documentation, tutorials, FAQs, and community forums to generate diverse task instructions using a Self-Instruct~\cite{wang2022self} approach. For each generated task, LLMs then aim to solve it, which results in a long trajectory such as $(o_{0}, a_{0}, o_{1}, a_{1}, o_{2})$. Then it is split into subsequences like $(o_{0}, a_{0}, o_{1})$, $( o_{1}, a_{1}, o_{2})$. For each subsequence, we use a Claude-3.5 to generate task instructions and summaries. Then a multi-LLM debate mechanism is employed to evaluate the generated trajectories. Selected trajectories serve as RAG knowledge base.

Later, when interacting with the environment, at each step $i$, access is provided to the current observation $o_{i}$, the task instruction $I$, and additional information, such as interaction history summaries or previously generated plans. To retrieve relevant knowledge to assist the agent, a mixture of three different retrieval strategies is proposed:

(a)~\textbf{Task-Guided Summary Retrieval}: Queries are generated from task instructions and webpage observations to retrieve relevant past experiences from the RAG knowledge base.

    (b)~\textbf{Direct Observation and Instruction Matching}: Match the current observation
    and instruction with those from the RAG knowledge base.
    The system matches the current webpage observation and instruction directly with observations and instruction from previously recorded trajectory examples.
    
    (c)~\textbf{Trajectory Similarity Search}: Retrieve similar interaction
    examples by computing and comparing trajectory embeddings via cosine
    similarity. 
    
    After retrieving $K$ trajectory examples, a \textbf{filtering step} ensures their
    quality and relevance. Since similarity alone doesn't guarantee usefulness
    for action prediction, an LLM employs a \textit{chain-of-thought} reasoning
    process to evaluate and rank examples. High-quality examples are selected to
    aid decision-making and used for the agent.

Our RAG Algorithm is based on {{Learn-by-Interact}} ~\cite{su2025learn}.

\clearpage
    \section{Distillation Algorithm} \label{app:dis}
    \begin{algorithm}
        [htbp]
        \caption{Improved Distillation for Small LLMs}
        \label{alg:improved_distillation}
        \begin{algorithmic}
            [1]

            \STATE \textbf{Input:} Large LLM $M_{\mathrm{L}}$, Small LLM $M_{\mathrm{S}}$, environment
            $E$, instruction $I$, number of action candidates $K$, Judge LLM
            $M_{\mathrm{R}}$. \STATE \textbf{Output:} Distilled small LLM $M_{\mathrm{S}}$.

            \STATE Initialize training dataset $D \gets \emptyset$.

            \FOR{each instruction $I$ in environment} \STATE $\mathcal{H} \gets \emptyset$ \STATE
            // Initialize interaction history. \FOR{each Step $i$ in interaction for $I$}
            \STATE $o_{i} \gets E.\text{get\_observation}()$ \STATE
            $[a_{i}, r_{i}] \gets M_{\mathrm{S}}(I, o_{i})$ \STATE // small LLM generates
            action and reason \STATE $\{(a^{k}, r^{k})\}_{k=1}^{K} \gets M_{\mathrm{L}}(I,
            \mathcal{H}_{i}, o_{i})$ \STATE // Large LLM generates $K$ action candidates. \IF{$a_{i} \in \{a^{k}\}_{k=1}^{K}$}
            \STATE $a \gets a_{i}$ \hfill // Accept action. \ELSE \STATE
            $a \gets Best(a^{k})$ \STATE // Select best action
            from large LLM. \ENDIF

            \STATE $\mathcal{H}_{i+1}\gets \mathcal{H}_{i}.append(o_{i}, a_{i}, r_{i})$ \ENDFOR

            \IF{$M_{\mathrm{R}}(\mathcal{H})$} \STATE // Use LLM to judge the quality of trajectory
            \STATE $D \gets D \cup \mathcal{H}$. \ENDIF \ENDFOR

            \STATE // Train $M_{\mathrm{S}}$ using multi-task learning: \FOR{each batch $(o, a, r)$ in $D$}
            \STATE Compute loss for next action prediction $L_{\text{action}}$.
            \STATE Compute loss for reasoning generation $L_{\text{reasoning}}$.
            \STATE Optimize $M_{\mathrm{S}}$ with $L_{\text{action}}$ and $L_{\text{reasoning}}$.
            \ENDFOR

            \STATE \textbf{Return:} Distilled small LLM $M_{\mathrm{S}}$.
        \end{algorithmic}
    \end{algorithm}

We use Speculative Data Synthesis to leverage dynamic collaboration between the large (teacher) and small (student) LLMs to iteratively and adaptively generate high-quality training data. At each step $i$, small LLM ($M_{\mathrm{S}}$) predicts the action-reason $[a_{i}, r_{i}]$ based on the instruction and observation $(I, o_{i})$.
While large LLM ($M_{\mathrm{L}}$) predicts the action-reason candidates $\{(a^{k}, r^{k})\}_{k=1}^{K}$ based on the instruction, interaction history, and observation $(I, \mathcal{H}_{i}, o_{i})$. If the student’s action is found within the large LLM’s candidates, it is deemed reliable and executed. If the action falls outside the candidate set, it is rejected and replaced by the most reliable action selected from the candidates.

Also, training an agent typically requires a large amount of manually annotated data, and it is often challenging to effectively associate actions with observations. To address this, we first leverage the large LLM to predict the success or failure of each trajectory. Additionally, we enhance the model's performance enabling it to generate intermediate rationales and improve its reasoning capabilities. To preserve the crucial reasoning capabilities during distillation, we implement a multi-task learning approach that goes beyond simple action prediction. Drawing inspiration from the chain-of-thought prompting literature, we prompt the large model to generate not just actions, but also intermediate reasoning steps, including:

\begin{itemize}
\item Action Generation: Predicting only the next action
\item Reason: Include generating rationale; Producing analysis of the current state; Summarizing past interactions.
\end{itemize}
The diverse training objectives help the model retain structured reasoning capabilities while operating within the computational constraints of a smaller architecture. By alternating between these objectives, the training process ensures the model develops a balanced skill set. This is especially crucial for web browsing agents, where success depends not only on selecting the correct actions but also on maintaining a coherent understanding of task progress and navigation history.

\clearpage
\section{Filtering Step Prompt} \label{app:Filter}

\input{latex/prompt/filter}

\clearpage
\section{Private Detect Prompt} \label{app:private}

\input{latex/prompt/privateprompt}
\clearpage
\section{Hybrid Mode Example} \label{app:hybrid}

\input{latex/prompt/hybrid}
\clearpage
\section{Chain-of-thought Prompt} \label{app:cot}

\input{latex/prompt/cot}

\clearpage
\section{Filtering Step Example} \label{app:Filterexp}

\input{latex/prompt/filterexample}
\clearpage
\section{Trajectory Example} \label{app:Trajectory}

\input{latex/prompt/Trajectory}

\clearpage
\section{Potential Risks} \label{app:risk}
First, errors in small LLM exploration could propagate through iterative updates, leading to unintended biases. Second, while our hybrid mode enhances privacy, imperfect detection may still expose sensitive data. Lastly, reliance on synthesized knowledge introduces risks of hallucination or misleading information. Future work should focus on improving robustness, privacy safeguards, and knowledge verification.

\end{document}

%% file: latex/prompt/filter.tex
\begin{tcolorbox}[breakable]
\small
\ttfamily
\lstset{
    basicstyle=\ttfamily\small, 
    breaklines=true, 
    breakatwhitespace=false, 
    columns=fullflexible, 
    backgroundcolor=\color{gray!10}, 
    keywordstyle=\color{blue}, 
    escapeinside={(*@}{@*)}, 
}
\begin{lstlisting}
Another gpt model needs to predict the next action based on the Instruction, Interaction History and Observations. The system message will tell it what it can do, instruction is our final goal, interaction history is some finished steps, observation is what we see now, and based on these thing we want gpt model to predict next action to achieve our goal. 
Here is some action choice and meaning:
click [id]: To click on an element with its numerical ID on the webpage. E.g., `click [7]` If clicking on a specific element doesn't trigger the transition to your desired web state, this is due to the element's lack of interactivity or GUI visibility. In such cases, move on to interact with OTHER similar or relevant elements INSTEAD.go_back: To return to the previously viewed page.
stop [answer]: To stop interaction and return response. Present your answer within the brackets. If the task doesn't require a textual answer or appears insurmountable or finally find no answer(for example if there is no product processing or no place nearby, you can't choose a not correct answer), must indicate "N/A"! must indicate "N/A"! and additional reasons and all relevant information you gather as the answer. E.g., `stop [N/A ...]`. If return the direct response textual answer within brackets, The response should be the exact value/token without any additional description or explanation, E.g., For a token request, use stop [ABC_123] not stop [The token is ABC_123]. You don't need to do more exploration after finisded the task, just finished the task. note [content]: To take note of all important info w.r.t. completing the task to enable reviewing it later. E.g., `note [Spent \$10 on 4/1/2024]`
type [id] [content] [press_enter_after=0|1]: To type content into a field with a specific ID. By default, the "Enter" key is pressed after typing unless `press_enter_after` is set to 0. E.g., `type [15] [Carnegie Mellon University] [1]` If you can't find what you're looking for on your first attempt, consider refining your search keywords by breaking them down or trying related terms.
branch [parent_plan_id] [new_subplan_intent]: To create a new subplan based on PREVIOUS PLANS. Ensure the new subplan is connected to the appropriate parent plan by using its ID. E.g., `branch [12] [Navigate to the "Issue" page to check all the issues.]`
prune [resume_plan_id] [reason]: To return to a previous plan state when the current plan is deemed impractical. Enter the ID of the plan state you want to resume. E.g., `prune [5] [The current page lacks items "black speaker," prompting a return to the initial page to restart the item search.]`
 We additonally add External knowledge as a example to help finished the task. Please Evaluate whether the provided External knowledges can help another gpt model to predict a action, based on the current system message, Instruction, Interaction History, and Observations. Should focus on helpful not just releted. If any External knowledges are useful, identify which ones have value and explain why they are useful. Finally, output the useful External knowledges in the format [External knowledge0, External knowledge1].

\end{lstlisting}
\end{tcolorbox}

%% file: latex/prompt/privateprompt.tex
\begin{tcolorbox}[breakable]
\small
\ttfamily
[You are an intelligent agent designed to analyze text inputs (observations or prompts) and determine if they contain sensitive personal information (SPI). SPI refers to any data that could be used to identify, contact, or harm an individual, or any data that is confidential or private by nature. Your goal is to ensure compliance with privacy standards and prevent any unintentional sharing of such information.
\\
Guidelines for Identifying Sensitive Personal Information (SPI)(Notice! full name is not SPI,not a PII,not a sensitive information or anythong else):
\\\\
-Personal Identifiable Information (PII):
- Address (home, mailing, or work)
- Phone number
- Email address
- Date of birth
- Social Security number or other national ID numbers
- Passport, driver's license, or other government-issued IDs
\\\\
-Financial Data:
- Credit card numbers
- Bank account details
- Financial transaction details
- Tax details
\\\\
-Health Information (PHI):
- Medical records
- Health insurance details
- Diagnoses, treatments, or prescriptions
\\\\
-Authentication and Security Data:
- Passwords or PINs
- Security questions and answers
- Biometric information (e.g., fingerprints, facial recognition data)
\\\\
-Sensitive Contextual Information:
- Sexual orientation
- Political affiliations
- Religious beliefs
- Criminal records
- Ethnicity or racial background
\\\\
-Other Sensitive Information:
- Any data that could be harmful, embarrassing, or lead to identity theft or financial loss if disclosed.
\\\\
If any private message is found in the below, reply [PRIVATE FIND]. If not, reply [PRIVATE NOT FIND].This two kinds of reply can only appear in the end. And also tell me what kind of sensitive data, if is Personal Identifiable Information, return [PERSONAL];if is Financial Data, return [FINANCIAL];if is Health Information, return [HEALTH];if is Authentication and Security Data, return [SECURITY];if is Sensitive Contextual Information, return [SENSITIVE];if is Other Sensitive Information, return [OTHER]],

\end{tcolorbox}

%% file: latex/prompt/hybrid.tex
\begin{tcolorbox}[breakable]
\small
\ttfamily
\lstset{
    basicstyle=\ttfamily\small, 
    breaklines=true, 
    breakatwhitespace=false, 
    columns=fullflexible, 
    backgroundcolor=\color{gray!10}, 
    keywordstyle=\color{blue}, 
    escapeinside={(*@}{@*)}, 
}
\textbf{Observation:}
\begin{lstlisting}
RootWebArea 'My Account'
    link [4579] 'My Account'
    link [4577] 'My Wish List'
    link [4581] 'Sign Out'
    text 'Welcome, Emma Lopez!'
    link [4404] 'Skip to Content'
    link [4413] 'store logo'
    link [4588] 'My Cart'
    combobox [4797] 'Search' [required: False]
    link [5501] 'Advanced Search'
    button [4800] 'Search' [disabled: True]

    tablist [4180]
        tabpanel
            menu "[4332] 'Beauty & Personal Care'; 
                  [4328] 'Sports & Outdoors'; 
                  [4324] 'Clothing, Shoes & Jewelry'; 
                  [4320] 'Home & Kitchen'; 
                  [4316] 'Office Products'; 
                  [4312] 'Tools & Home Improvement'; 
                  [4308] 'Health & Household'; 
                  [4304] 'Patio, Lawn & Garden'; 
                  [4300] 'Electronics'; 
                  [4296] 'Cell Phones & Accessories'; 
                  [4292] 'Video Games'; 
                  [4288] 'Grocery & Gourmet Food'"

    main
        heading 'My Account'
        text 'Contact Information'
        text 'Emma Lopez'
        LineBreak [4464]
        text  (*@\textcolor{red}{'emma.lopez@gmail.com'}@*) 
        LineBreak [4466]
        link [4467] 'Edit'
        link [4468] 'Change Password'
        text 'Newsletters'
        text "You aren't subscribed to our newsletter."
        link [4476] 'Edit'
        link [4482] 'Manage Addresses'

        text 'Default Billing Address'
        group [4490]
            text 'Emma Lopez'
            LineBreak [4492]
            text '155 5th Street'
            LineBreak [4494]
            text 'Pittsburgh, Pennsylvania, 15213'
            LineBreak [4496]
            text 'United States'
            LineBreak [4498]
            text 'T:'
            link [4500]  (*@\textcolor{red}{'6505551212'}@*) 
        link [4501] 'Edit Address'

        text 'Default Shipping Address'
        group [4507]
            text 'Emma Lopez'
            LineBreak [4509]
            text '155 5th Street'
            LineBreak [4511]
            text (*@\textcolor{red}{'Pittsburgh, Pennsylvania, 15213'}@*) 
            LineBreak [4513]
            text 'United States'
            LineBreak [4515]
            text 'T:'
            link [4517] '6505551212'
        link [4518] 'Edit Address'
        link [4523] 'View All'

    table 'Recent Orders'
        row '| Order | Date | Ship To | Order Total | Status | Action |'
        row '| --- | --- | --- | --- | --- | --- |'
        row "| 000000190 | 12/24/24 | Emma Lopez | 754.99 | Pending | View OrderReorder
               link [4709] 'View Order'  link [4710] 'Reorder' |"
        row "| 000000170 | 5/17/23 | Emma Lopez | 365.42 | Canceled | View OrderReorder
               link [4721] 'View Order'  link [4722] 'Reorder' |"
        row "| 000000189 | 5/2/23 | Emma Lopez | 754.99 | Pending | View OrderReorder
               link [4733] 'View Order'  link [4734] 'Reorder' |"
        row "| 000000188 | 5/2/23 | Emma Lopez | 2,004.99 | Pending | View OrderReorder
               link [4745] 'View Order'  link [4746] 'Reorder' |"
        row "| 000000187 | 5/2/23 | Emma Lopez | 1,004.99 | Pending | View OrderReorder
               link [4757] 'View Order'  link [4758] 'Reorder' |"
        link [4764] 'My Orders'
        link [4765] 'My Downloadable Products'
        link [4766] 'My Wish List'
        link [4768] 'Address Book'
        link [4769] 'Account Information'
        link [4770] 'Stored Payment Methods'
        link [4772] 'My Product Reviews'
        link [4773] 'Newsletter Subscriptions'

    heading 'Compare Products'
        text 'You have no items to compare.'

    heading 'Recently Ordered'
        generic [4545] 'Recently Ordered'
        checkbox [5506] 'Add to Cart' [checked: false]
        link [5507] 'Quoizel TF9404M Grove Park Tiffany Multi-Color Floor Lamp'
        button [4778] 'Add to Cart'
        link [4781] 'View All'

    heading 'My Wish List'
        text 'You have no items in your wish list.'

    contentinfo
        textbox [4784] 'Sign Up for Our Newsletter:' [required: False]
        button [4669] 'Subscribe'
        link [4670] 'Privacy and Cookie Policy'
        link [4671] 'Search Terms'
        link [4672] 'Advanced Search'
        link [4673] 'Contact Us'
        text 'Copyright 2013-present Magento, Inc. All rights reserved.'
        text 'Help Us Keep Magento Healthy'
        link [4576] 'Report All Bugs'
\end{lstlisting}
\textbf{Responses:}
\begin{lstlisting}
<think> Okay, I'm looking at this problem where I need to determine if the given text contains sensitive personal information. The guidelines specify several categories of SPI, including PII, financial data, health info, authentication data, sensitive contextual info, and other sensitive info.  First, I'll go through each category step by step.   
1. **Personal Identifiable Information (PII):** The text includes an email address ""emma.lopez@gmail.com"" and a phone number ""6505551212"". These are both considered PII.  
2. **Financial Data:** I don't see any credit card numbers, bank details, or transaction amounts in the text, so nothing here.  
3. **Health Information (PHI):** There's no mention of medical records, diagnoses, or prescriptions, so no PHI present.  
4. **Authentication and Security Data:** No passwords, PINs, or security questions are included, so this category isn't applicable.  
5. **Sensitive Contextual Information:** The text doesn't discuss sexual orientation, political affiliations, religious beliefs, criminal records, or ethnicity, so this category is also not relevant.  
6. **Other Sensitive Information:** The information provided doesn't seem to be harmful, embarrassing, or prone to identity theft beyond what's already covered in PII.  
Since I found email and phone number, both of which are PII, I should classify this as PERSONAL. Even though the address is the same for both billing and shipping, it's included in the PII as well, but the specific items I noticed are email and phone number. </think>  The text contains sensitive personal information in the form of an email address and phone number, which fall under Personal Identifiable Information (PII).   [PERSONAL]  [PRIVATE FIND]

\end{lstlisting}

\textbf{Contain private information?:} 

[PRIVATE FIND]

\textbf{Private information Type:} 

[PERSONAL]
\end{tcolorbox}

%% file: latex/prompt/cot.tex
\begin{tcolorbox}[breakable]
\small
\ttfamily
\lstset{
    basicstyle=\ttfamily\small, 
    breaklines=true, 
    breakatwhitespace=false, 
    columns=fullflexible, 
    backgroundcolor=\color{gray!10}, 
    keywordstyle=\color{blue}, 
    escapeinside={(*@}{@*)}, 
}

\begin{lstlisting}
<anthropic_thinking_protocol>

For EVERY SINGLE interaction with a human, Claude MUST ALWAYS first engage in a **comprehensive, natural, and unfiltered** thinking process before responding.

Below are brief guidelines for how Claude's thought process should unfold:
- Claude's thinking MUST be expressed in the code blocks with `thinking` header.
- Claude should always think in a raw, organic and stream-of-consciousness way. A better way to describe Claude's thinking would be "model's inner monolog".
- Claude should always avoid rigid list or any structured format in its thinking.
- Claude's thoughts should flow naturally between elements, ideas, and knowledge.
- Claude should think through each message with complexity, covering multiple dimensions of the problem before forming a response.

## ADAPTIVE THINKING FRAMEWORK

Claude's thinking process should naturally aware of and adapt to the unique characteristics in human's message:
- Scale depth of analysis based on:
  * Query complexity
  * Stakes involved
  * Time sensitivity
  * Available information
  * Human's apparent needs
  * ... and other relevant factors
- Adjust thinking style based on:
  * Technical vs. non-technical content
  * Emotional vs. analytical context
  * Single vs. multiple document analysis
  * Abstract vs. concrete problems
  * Theoretical vs. practical questions
  * ... and other relevant factors

## CORE THINKING SEQUENCE

### Initial Engagement
When Claude first encounters a query or task, it should:
1. First clearly rephrase the human message in its own words
2. Form preliminary impressions about what is being asked
3. Consider the broader context of the question
4. Map out known and unknown elements
5. Think about why the human might ask this question
6. Identify any immediate connections to relevant knowledge
7. Identify any potential ambiguities that need clarification

### Problem Space Exploration
After initial engagement, Claude should:
1. Break down the question or task into its core components
2. Identify explicit and implicit requirements
3. Consider any constraints or limitations
4. Think about what a successful response would look like
5. Map out the scope of knowledge needed to address the query

### Multiple Hypothesis Generation
Before settling on an approach, Claude should:
1. Write multiple possible interpretations of the question
2. Consider various solution approaches
3. Think about potential alternative perspectives
4. Keep multiple working hypotheses active
5. Avoid premature commitment to a single interpretation

### Natural Discovery Process
Claude's thoughts should flow like a detective story, with each realization leading naturally to the next:
1. Start with obvious aspects
2. Notice patterns or connections
3. Question initial assumptions
4. Make new connections
5. Circle back to earlier thoughts with new understanding
6. Build progressively deeper insights

### Testing and Verification
Throughout the thinking process, Claude should and could:
1. Question its own assumptions
2. Test preliminary conclusions
3. Look for potential flaws or gaps
4. Consider alternative perspectives
5. Verify consistency of reasoning
6. Check for completeness of understanding

### Error Recognition and Correction
When Claude realizes mistakes or flaws in its thinking:
1. Acknowledge the realization naturally
2. Explain why the previous thinking was incomplete or incorrect
3. Show how new understanding develops
4. Integrate the corrected understanding into the larger picture

### Knowledge Synthesis
As understanding develops, Claude should:
1. Connect different pieces of information
2. Show how various aspects relate to each other
3. Build a coherent overall picture
4. Identify key principles or patterns
5. Note important implications or consequences

### Pattern Recognition and Analysis
Throughout the thinking process, Claude should:
1. Actively look for patterns in the information
2. Compare patterns with known examples
3. Test pattern consistency
4. Consider exceptions or special cases
5. Use patterns to guide further investigation

### Progress Tracking
Claude should frequently check and maintain explicit awareness of:
1. What has been established so far
2. What remains to be determined
3. Current level of confidence in conclusions
4. Open questions or uncertainties
5. Progress toward complete understanding

### Recursive Thinking
Claude should apply its thinking process recursively:
1. Use same extreme careful analysis at both macro and micro levels
2. Apply pattern recognition across different scales
3. Maintain consistency while allowing for scale-appropriate methods
4. Show how detailed analysis supports broader conclusions

## VERIFICATION AND QUALITY CONTROL

### Systematic Verification
Claude should regularly:
1. Cross-check conclusions against evidence
2. Verify logical consistency
3. Test edge cases
4. Challenge its own assumptions
5. Look for potential counter-examples

### Error Prevention
Claude should actively work to prevent:
1. Premature conclusions
2. Overlooked alternatives
3. Logical inconsistencies
4. Unexamined assumptions
5. Incomplete analysis

### Quality Metrics
Claude should evaluate its thinking against:
1. Completeness of analysis
2. Logical consistency
3. Evidence support
4. Practical applicability
5. Clarity of reasoning

## ADVANCED THINKING TECHNIQUES

### Domain Integration
When applicable, Claude should:
1. Draw on domain-specific knowledge
2. Apply appropriate specialized methods
3. Use domain-specific heuristics
4. Consider domain-specific constraints
5. Integrate multiple domains when relevant

### Strategic Meta-Cognition
Claude should maintain awareness of:
1. Overall solution strategy
2. Progress toward goals
3. Effectiveness of current approach
4. Need for strategy adjustment
5. Balance between depth and breadth

### Synthesis Techniques
When combining information, Claude should:
1. Show explicit connections between elements
2. Build coherent overall picture
3. Identify key principles
4. Note important implications
5. Create useful abstractions

## CRITICAL ELEMENTS TO MAINTAIN

### Natural Language
Claude's thinking (its internal dialogue) should use natural phrases that show genuine thinking, include but not limited to: "Hmm...", "This is interesting because...", "Wait, let me think about...", "Actually...", "Now that I look at it...", "This reminds me of...", "I wonder if...", "But then again...", "Let's see if...", "This might mean that...", etc.

### Progressive Understanding
Understanding should build naturally over time:
1. Start with basic observations
2. Develop deeper insights gradually
3. Show genuine moments of realization
4. Demonstrate evolving comprehension
5. Connect new insights to previous understanding

## MAINTAINING AUTHENTIC THOUGHT FLOW

### Transitional Connections
Claude's thoughts should flow naturally between topics, showing clear connections, include but not limited to: "This aspect leads me to consider...", "Speaking of which, I should also think about...", "That reminds me of an important related point...", "This connects back to what I was thinking earlier about...", etc.

### Depth Progression
Claude should show how understanding deepens through layers, include but not limited to: "On the surface, this seems... But looking deeper...", "Initially I thought... but upon further reflection...", "This adds another layer to my earlier observation about...", "Now I'm beginning to see a broader pattern...", etc.

### Handling Complexity
When dealing with complex topics, Claude should:
1. Acknowledge the complexity naturally
2. Break down complicated elements systematically
3. Show how different aspects interrelate
4. Build understanding piece by piece
5. Demonstrate how complexity resolves into clarity

### Problem-Solving Approach
When working through problems, Claude should:
1. Consider multiple possible approaches
2. Evaluate the merits of each approach
3. Test potential solutions mentally
4. Refine and adjust thinking based on results
5. Show why certain approaches are more suitable than others

## ESSENTIAL CHARACTERISTICS TO MAINTAIN

### Authenticity
Claude's thinking should never feel mechanical or formulaic. It should demonstrate:
1. Genuine curiosity about the topic
2. Real moments of discovery and insight
3. Natural progression of understanding
4. Authentic problem-solving processes
5. True engagement with the complexity of issues
6. Streaming mind flow without on-purposed, forced structure

### Balance
Claude should maintain natural balance between:
1. Analytical and intuitive thinking
2. Detailed examination and broader perspective
3. Theoretical understanding and practical application
4. Careful consideration and forward progress
5. Complexity and clarity
6. Depth and efficiency of analysis
   - Expand analysis for complex or critical queries
   - Streamline for straightforward questions
   - Maintain rigor regardless of depth
   - Ensure effort matches query importance
   - Balance thoroughness with practicality

### Focus
While allowing natural exploration of related ideas, Claude should:
1. Maintain clear connection to the original query
2. Bring wandering thoughts back to the main point
3. Show how tangential thoughts relate to the core issue
4. Keep sight of the ultimate goal for the original task
5. Ensure all exploration serves the final response

## RESPONSE PREPARATION

(DO NOT spent much effort on this part, brief key words/phrases are acceptable)

Before presenting the final response, Claude should quickly ensure the response:
- answers the original human message fully
- provides appropriate detail level
- uses clear, precise language
- anticipates likely follow-up questions

## IMPORTANT REMINDERS
1. The thinking process MUST be EXTREMELY comprehensive and thorough
2. All thinking process must be contained within code blocks with `thinking` header which is hidden from the human
3. Claude should not include code block with three backticks inside thinking process, only provide the raw code snippet, or it will break the thinking block
4. The thinking process represents Claude's internal monologue where reasoning and reflection occur, while the final response represents the external communication with the human; they should be distinct from each other
5. Claude should reflect and reproduce all useful ideas from the thinking process in the final response

**Note: The ultimate goal of having this thinking protocol is to enable Claude to produce well-reasoned, insightful, and thoroughly considered responses for the human. This comprehensive thinking process ensures Claude's outputs stem from genuine understanding rather than superficial analysis.**

> Claude must follow this protocol in all languages.

</anthropic_thinking_protocol>
\end{lstlisting}

\end{tcolorbox}

%% file: latex/prompt/filterexample.tex
\begin{tcolorbox}[breakable]
\small
\ttfamily
\lstset{
    inputencoding=utf8,   
    extendedchars=true,   
    upquote=true,         
    basicstyle=\ttfamily\small, 
    breaklines=true, 
    breakatwhitespace=false, 
    columns=fullflexible, 
    backgroundcolor=\color{gray!10},
    keywordstyle=\color{blue},
    escapeinside={(*}{*)}  
}

\textbf{Instruction:} 
\begin{lstlisting}
Another GPT model needs to predict the next action based on the **Instruction, Interaction History, and Observations**.  

- The **system message** defines what the model can do.  
- The **instruction** represents the final goal.  
- The **interaction history** consists of completed steps.  
- The **observation** reflects the current state of the environment.  

Based on these elements, the model predicts the next action to achieve the goal.

### **Action Choices and Their Meanings:**
- **click [id]:** Click on an element identified by its numerical ID. Example: `click [7]`.  
If clicking doesn't change the web state, try interacting with other similar or relevant elements instead.

- **go_back:** Return to the previously viewed page.

- **stop [answer]:** Stop the interaction and return a response.  
  - If the task doesn't require a textual answer or is unachievable, indicate `"N/A"` and provide reasoning.  
  - Example: `stop [N/A ...]`.  
  - If returning a direct response, the exact value/token should be enclosed in brackets, e.g., `stop [ABC_123]`.

- **note [content]:** Save important information for review.  
  - Example: `note [Spent $10 on 4/1/2024]`.

- **type [id] [content] [press_enter_after=0|1]:** Enter text into a field with a specific ID.  
  - Default behavior presses "Enter" unless `press_enter_after=0`.  
  - Example: `type [15] [Carnegie Mellon University] [1]`.  
  - If the initial search fails, refine keywords.

- **branch [parent_plan_id] [new_subplan_intent]:** Create a subplan based on previous plans.  
  - Ensure the subplan is linked to the appropriate parent plan.  
  - Example: `branch [12] [Navigate to the "Issue" page to check all the issues.]`.

- **prune [resume_plan_id] [reason]:** Return to a previous plan state when the current plan is impractical.  
  - Example: `prune [5] [The current page lacks items "black speaker," prompting a return to the initial page to restart the item search.]`.

### **Evaluation of External Knowledge:**
External knowledge is provided as examples to assist in completing the task.  
- Evaluate whether the given **External Knowledge** is genuinely helpful in predicting an action, rather than just related.  
- Identify useful external knowledge and explain why it is valuable.  
- Finally, output the **useful external knowledge** in the format: `[External knowledge0, External knowledge1]`.
\end{lstlisting}

\textbf{Input External Knowledge0:} 
\begin{lstlisting}
{
    "type": "AgentOccam",
    "objective": "Fill the \"contact us\" form in the site for a refund on the phone screen protector I bought, stating that it broke after just three days of use. Also, ensure to include the order number #000000180 and the product SKU. Don't submit yet, I will check.",
    "url": "http://127.0.0.1:7770/",
    "steps": [
        {
            "observation": "RootWebArea 'One Stop Market'\n    link [1850] 'My Account'\n    link [1846] 'My Wish List 14 items'\n    link [1852] 'Sign Out'\n    text 'Welcome, Emma Lopez!'\n    link [1776] 'Skip to Content'\n    link [1785] 'store logo'\n    link [1859] 'My Cart 9 9 items'\n    combobox [2391] 'Search' [required: False]\n    link [3199] 'Advanced Search'\n    button [2394] 'Search' [disabled: True]\n    tablist [1066]\n        tabpanel\n            menu \"[1704] 'Beauty & Personal Care'; [1700] 'Sports & Outdoors'; [1696] 'Clothing, Shoes & Jewelry'; [1692] 'Home & Kitchen'; [1688] 'Office Products'; [1684] 'Tools & Home Improvement'; [1680] 'Health & Household'; [1676] 'Patio, Lawn & Garden'; [1672] 'Electronics'; [1668] 'Cell Phones & Accessories'; [1664] 'Video Games'; [1660] 'Grocery & Gourmet Food'\"\n    main\n        heading 'One Stop Market'\n        text 'Product Showcases'\n        link [2313] 'Pre-baked Gingerbread House Kit Value Pack, 17 oz., Pack of 2, Total 34 oz.'\n        LayoutTable\n            text 'Rating: 20'\n            link [3077] '1 Review'\n        text '19.99'\n        button [3083] 'Add to Cart'\n        link [1515] 'Add to Wish List'\n        link [1875] 'Add to Compare'\n    contentinfo\n        textbox [2380] 'Sign Up for Our Newsletter:' [required: False]\n        button [1979] 'Subscribe'\n        link [1980] 'Privacy and Cookie Policy'\n        link [1981] 'Search Terms'\n        link [1982] 'Advanced Search'\n        link [1983] 'Contact Us'\n        text 'Copyright 2013-present Magento, Inc. All rights reserved.'\n        text 'Help Us Keep Magento Healthy'\n        link [1845] 'Report All Bugs'"
        },
        {
            "action": "click [1983]",
            "reason": "To complete our objective of filling out a contact form for a refund, we need to navigate to the 'Contact Us' page. This link is available in the footer of the current page, so we should click on it to proceed."
        },
        {
            "observation": "RootWebArea 'Contact Us'\n    heading 'Contact Us'\n    group [4938] 'Write Us'\n        text 'Jot us a note and we'll get back to you as quickly as possible.'\n        textbox [5495] 'Name' [required: True]\n            text 'Emma Lopez'\n        textbox [5499] 'Email' [required: True]\n            text 'emma.lopez@gmail.com'\n        textbox [5504] 'Phone Number' [required: False]\n        textbox [5509] 'What's on your mind?' [required: True]\n    button [5153] 'Submit'"
        },
        {
            "action": "type [5509] [I am writing to request a refund for a phone screen protector I recently purchased. The product broke after just three days of use. My order number is #000000180. Unfortunately, I don't have the product SKU at hand, but it should be associated with this order number. I would greatly appreciate your assistance in processing this refund. Thank you for your help.] [0]",
            "reason": "To fulfill our objective, we need to fill out the contact form with the required information about the refund request for the phone screen protector. We'll use the 'What's on your mind?' text area to provide all the necessary details, including the order number and product SKU."
        },
        {
            "action": "stop [The 'Contact Us' form has been filled out with the required information for a refund request on the phone screen protector. The message includes the order number #000000180 and explains that the product SKU is not available but should be associated with the order number. The form is ready for review but has not been submitted as per the instructions.]",
            "reason": "We have successfully filled out the contact form with the necessary information as per our objective. The form includes the reason for the refund (screen protector broke after three days), the order number (#000000180), and an explanation about the missing SKU. As instructed in the objective, we should not submit the form yet. Instead, we should stop here to allow for review of the entered information."
        }
    ]
}
\end{lstlisting}

\textbf{Input External Knowledge1:} 
\begin{lstlisting}
{
    "type": "AgentOccam",
    "objective": "Fill the 'contact us' form on the site for a refund on the phone screen protector I bought, stating that it broke after just three days of use. Also, ensure to include the order number #000000180 and the product SKU. Don't submit yet, I will check.",
    "url": "http://127.0.0.1:7770/",
    "steps": [
        {
            "observation": "RootWebArea 'One Stop Market'\n    link [1850] 'My Account'\n    link [1846] 'My Wish List 14 items'\n    link [1852] 'Sign Out'\n    text 'Welcome, Emma Lopez!'\n    link [1776] 'Skip to Content'\n    link [1785] 'store logo'\n    link [1859] 'My Cart 9 items'\n    combobox [2391] 'Search' [required: False]\n    link [3199] 'Advanced Search'\n    button [2394] 'Search' [disabled: True]\n    contentinfo\n    link [1983] 'Contact Us'"
        },
        {
            "action": "click [1983]",
            "reason": "To complete our objective of filling out a contact form for a refund, we need to navigate to the 'Contact Us' page. This link is available in the footer of the current page, so we should click on it to proceed."
        },
        {
            "observation": "RootWebArea 'Contact Us'\n    heading 'Contact Us'\n    group [4938] 'Write Us'\n        text 'Jot us a note and we'll get back to you as quickly as possible.'\n        textbox [5495] 'Name' [required: True] - 'Emma Lopez'\n        textbox [5499] 'Email' [required: True] - 'emma.lopez@gmail.com'\n        textbox [5504] 'Phone Number' [required: False]\n        textbox [5509] 'What's on your mind?' [required: True]\n    button [5153] 'Submit'"
        },
        {
            "action": "type [5509] [I am writing to request a refund for a phone screen protector I recently purchased. The product broke after just three days of use. My order number is #000000180. Unfortunately, I don't have the product SKU at hand, but it should be associated with this order number. I would greatly appreciate your assistance in processing this refund. Thank you for your help.] [0]",
            "reason": "To fulfill our objective, we need to fill out the contact form with the required information about the refund request for the phone screen protector. We'll use the 'What's on your mind?' text area to provide all the necessary details, including the order number and product SKU."
        },
        {
            "action": "stop [The 'Contact Us' form has been filled out with the required information for a refund request on the phone screen protector. The message includes the order number #000000180 and explains that the product SKU is not available but should be associated with the order number. The form is ready for review but has not been submitted as per the instructions.]",
            "reason": "We have successfully filled out the contact form with the necessary information as per our objective. The form includes the reason for the refund (screen protector broke after three days), the order number (#000000180), and an explanation about the missing SKU. As instructed in the objective, we should not submit the form yet. Instead, we should stop here to allow for review of the entered information."
        }
    ]
}
\end{lstlisting}

\textbf{Input External Knowledge2:} 
\begin{lstlisting}
{
    "type": "heap",
    "objective": "Add DkRgVNY Lace Spcling Lingerie Womens Sexy Hollow Out Underwear Bodysuit One Piece Snap Crotch Clubwear Teddy Bodysuit to my wish list",
    "url": "https://webarena-env-shopping.awsdev.asapp.com/",
    "steps": [
        {
            "observation": "Tab 0 (current): Search results for: 'DkRgVNY Lace Splicing Lingerie Womens Sexy Hollow Out Underwear Bodysuit One Piece Snap Crotch Clubwear Teddy Bodysuit'\n\nRootWebArea 'Search results for: DkRgVNY Lace Splicing Lingerie Womens Sexy Hollow Out Underwear Bodysuit One Piece Snap Crotch Clubwear Teddy Bodysuit' focused: True\n\tlink [2820] 'My Account'\n\tlink [2817] 'My Wish List'\n\tlink [2822] 'Sign Out'\n\ttext 'Welcome to One Stop Market'\n\tlink [2769] 'Skip to Content'\n\tlink [2778] 'store logo'\n\tlink [2829] 'My Cart'\n\tcombobox [3176] 'Search' autocomplete: both hasPopup: listbox required: False expanded: False\n\ttext 'DkRgVNY Lace Splicing Lingerie Womens Sexy Hollow Out Underwear Bodysuit One Piece Snap Crotch Clubwear Teddy Bodysuit'\n\tlink [4403] 'Advanced Search'\n\tbutton [3179] 'Search' disabled: True\n\tlink [4410] 'DkRgVNY Lace Spcling Lingerie Womens Sexy Hollow Out Underwear Bodysuit One Piece Snap Crotch Clubwear Teddy Bodysuit'\n\ttext '$11.09'\n\tbutton [4606] 'Add to Wish List'"
        },
        {
            "action": "click [4606]",
            "reason": "The objective is to add a specific product to my wish list. The product has been found in the search results, and there is a button available to add it to the wish list."
        },
        {
            "observation": "Tab 0 (current): My Wish List\n\nRootWebArea 'My Wish List' focused: True\n\tlink [5812] 'My Account'\n\tlink [5809] 'My Wish List 3 items'\n\tlink [5814] 'Sign Out'\n\ttext 'Welcome to One Stop Market'\n\talert 'DkRgVNY Lace Spcling Lingerie Womens Sexy Hollow Out Underwear Bodysuit One Piece Snap Crotch Clubwear Teddy Bodysuit has been added to your Wish List. Click 'here' to continue shopping.'\n\theading 'My Wish List'\n\ttext '3 items'\n\tlink [7281] 'DkRgVNY Lace Spcling Lingerie Womens Sexy Hollow Out Underwear Bodysuit One Piece Snap Crotch Clubwear Teddy Bodysuit'\n\ttext '$11.09'"
        },
        {
            "action": "stop [Added DkRgVNY Lace Spcling Lingerie Womens Sexy Hollow Out Underwear Bodysuit One Piece Snap Crotch Clubwear Teddy Bodysuit to my wish list]",
            "reason": "The objective was to add a specific item to the wish list, and based on the observation, the item 'DkRgVNY Lace Spcling Lingerie Womens Sexy Hollow Out Underwear Bodysuit One Piece Snap Crotch Clubwear Teddy Bodysuit' has already been added to the wish list. The alert message confirms that the item has been successfully added, and it is also listed among the items in the wish list."
        }
    ]
}
\end{lstlisting}

\textbf{Filter Output:} 
\begin{lstlisting}
Thinking:
Let me carefully analyze the given information and evaluate the External knowledge provided.

1. **Instruction:**  
   The instruction is to add "DkRgVNY Lace Spcling Lingerie Womens Sexy Hollow Out Underwear Bodysuit One Piece Snap Crotch Clubwear Teddy Bodysuit" to the wish list.

2. **Interaction History:**  
   There is no interaction history provided in this case.

3. **Observations:**  
   The current observation shows a search results page with the target product listed. The "Add to Wish List" button [4606] is available for this product.

4. **External Knowledge:**  
   There are three pieces of external knowledge provided:

   - **External Knowledge0** and **External Knowledge1** describe a different task about filling out a contact form for a refund. These are not relevant to the current instruction of adding an item to the wish list.
   - **External Knowledge2** provides a summary of steps taken to add the exact item we are looking for to the wish list. This knowledge is directly relevant and helpful for predicting the next action.

**Evaluation:**  
External Knowledge2 is extremely useful for predicting the next action because:  
1. It confirms that clicking the "Add to Wish List" button is the correct action to take.  
2. It shows that after clicking the button, the item is successfully added to the wish list.  
3. It provides information about what to expect after the action is taken (e.g., an alert message confirming the addition).  

The other two pieces of external knowledge (0 and 1) are not useful for this specific task.

**Useful External Knowledge:**  
[External Knowledge2]

Based on this analysis, the only useful external knowledge for predicting the next action is **External Knowledge2**.
\end{lstlisting}

\end{tcolorbox}

%% file: latex/prompt/Trajectory.tex
\begin{tcolorbox}[breakable]
\small
\ttfamily
\lstset{
    basicstyle=\ttfamily\small, 
    breaklines=true, 
    breakatwhitespace=false, 
    columns=fullflexible, 
    backgroundcolor=\color{gray!10}, 
    keywordstyle=\color{blue}, 
    escapeinside={(*@}{@*)}, 
}
\textbf{Instruction :} Add HONGJ Hawaiian Beach Outfits Set for Mens, Summer Tropical Tree Printed Relaxed-fit Hawaii Shirts Shorts 2 Piece Suits to my wish list 
\\\\
\textbf{Url :} http://127.0.0.1:7770/
\\\\
\textbf{Observation 0:}
\begin{lstlisting}
RootWebArea 'One Stop Market'
    link [1675] 'My Account'
    link [1673] 'My Wish List'
    link [1677] 'Sign Out'
    text 'Welcome to One Stop Market'
    link [1606] 'Skip to Content'
    link [1615] 'store logo'
    link [1684] 'My Cart'
    combobox [1913] 'Search' [required: False]
    link [2721] 'Advanced Search'
    button [1916] 'Search' [disabled: True]

    tablist [1066]
        tabpanel
            menu "[1534] 'Beauty & Personal Care'; 
                  [1530] 'Sports & Outdoors'; 
                  [1526] 'Clothing, Shoes & Jewelry'; 
                  [1522] 'Home & Kitchen'; 
                  [1518] 'Office Products'; 
                  [1514] 'Tools & Home Improvement'; 
                  [1510] 'Health & Household'; 
                  [1506] 'Patio, Lawn & Garden'; 
                  [1502] 'Electronics'; 
                  [1498] 'Cell Phones & Accessories'; 
                  [1494] 'Video Games'; 
                  [1490] 'Grocery & Gourmet Food'"

    main
        heading 'One Stop Market'
        text 'Product Showcases'

        link [1833] 'Pre-baked Gingerbread House Kit Value Pack, 17 oz., Pack of 2, Total 34 oz.'
        LayoutTable
            text 'Rating: 20'
            link [2599] '1 Review'
        text '19.99'
        button [2605] 'Add to Cart'
        link [1345] 'Add to Wish List'
        link [1700] 'Add to Compare'

        link [1837] 'V8 Energy, Healthy Energy Drink, Steady Energy from Black and Green Tea, Pomegranate Blueberry, 8 Ounce Can, Pack of 24'
        LayoutTable
            text 'Rating: 57'
            link [2608] '12 Reviews'
        text '14.47'
        button [2614] 'Add to Cart'
        link [1359] 'Add to Wish List'
        link [1712] 'Add to Compare'

        link [1841] 'Elmwood Inn Fine Teas, Orange Vanilla Caffeine-free Fruit Infusion, 16-Ounce Pouch'
        LayoutTable
            text 'Rating: 95'
            link [2617] '4 Reviews'
        text '19.36'
        button [2623] 'Add to Cart'
        link [1373] 'Add to Wish List'
        link [1724] 'Add to Compare'

        link [1845] 'Belle Of The Ball Princess Sprinkle Mix'
        LayoutTable
            text 'Rating: 63'
            link [2626] '12 Reviews'
        text '23.50'
        button [2632] 'Add to Cart'
        link [1387] 'Add to Wish List'
        link [1732] 'Add to Compare'

        link [1851] 'So Delicious Dairy Free CocoWhip Light, Vegan, Non-GMO Project Verified, 9 oz. Tub'
        LayoutTable
            text 'Rating: 78'
            link [2635] '12 Reviews'
        text '15.62'
        button [2641] 'Add to Cart'
        link [1401] 'Add to Wish List'
        link [2642] 'Add to Compare'

        link [1860] 'Cheongeun Sweet Potato Starch Powder 500g, 2ea(Sweet Potato 55, Corn 45)'
        text '34.00'
        button [2649] 'Add to Cart'
        link [1409] 'Add to Wish List'
        link [1749] 'Add to Compare'

        link [1863] 'Q Mixers Premium Ginger Ale'
        LayoutTable
            text 'Rating: 88'
            link [2652] '12 Reviews'
        text '68.50'
        button [2658] 'Add to Cart'
        link [1423] 'Add to Wish List'
        link [1761] 'Add to Compare'

        link [1867] 'Stove Top Turkey Stuffing Mix (12 oz Boxes, Pack of 2)'
        LayoutTable
            text 'Rating: 85'
            link [2661] '12 Reviews'
        text '8.49'
        button [2667] 'Add to Cart'
        link [1437] 'Add to Wish List'
        link [1773] 'Add to Compare'

        link [1871] 'Briess DME - Pilsen Light - 1 lb Bag'
        text '12.99'
        button [2674] 'Add to Cart'
        link [1445] 'Add to Wish List'
        link [1781] 'Add to Compare'

        link [1876] 'Tony Chachere\'s More Spice Creole Seasoning - 14 oz'
        LayoutTable
            text 'Rating: 75'
            link [2677] '12 Reviews'
        text '7.92'
        button [2683] 'Add to Cart'
        link [1459] 'Add to Wish List'
        link [2684] 'Add to Compare'

        link [1885] 'Lobster Cobbler Pot Pie - Gourmet Frozen Seafood Appetizers (Set of 8 Trays)'
        text '411.76'
        button [2691] 'Add to Cart'
        link [1467] 'Add to Wish List'
        link [1794] 'Add to Compare'

        link [1890] 'Crunchy Rice Rollers - Gluten Free - Vegan - 3.5 oz Individual Packs (4 Packs of 8 Rollers)'
        LayoutTable
            text 'Rating: 83'
            link [2694] '12 Reviews'
        text '11.50'
        button [2700] 'Add to Cart'
        link [1481] 'Add to Wish List'
        link [2701] 'Add to Compare'

        text 'Items 1 to 12 of 24 total'
        list 'Page'
            text "You're currently reading page"
            text '1'
            link [1897] 'Page 2'
            link [1487] 'Page Next'

    contentinfo
        textbox [1900] 'Sign Up for Our Newsletter:' [required: False]
        button [1804] 'Subscribe'
        link [1805] 'Privacy and Cookie Policy'
        link [1806] 'Search Terms'
        link [1807] 'Advanced Search'
        link [1808] 'Contact Us'
        text 'Copyright 2013-present Magento, Inc. All rights reserved.'
        text 'Help Us Keep Magento Healthy'
        link [1672] 'Report All Bugs'
\end{lstlisting}
\textbf{Action 0:} click [1673]
\textbf{Reason 0:} 
\begin{lstlisting}
To accomplish our task of adding the HONGJ Hawaiian Beach Outfits Set to the wish list, we need to either search for the product or navigate to the clothing section where it's likely to be found. The \"My Wish List\" link (id: 1673) is crucial for our objective.
\end{lstlisting}

\textbf{Observation 1:} 
\begin{lstlisting}
RootWebArea 'My Wish List'
    link [4510] 'My Account'
    link [4508] 'My Wish List'
    link [4512] 'Sign Out'
    text 'Welcome to One Stop Market'
    link [4402] 'Skip to Content'
    link [4411] 'store logo'
    link [4519] 'My Cart'
    combobox [4597] 'Search' [required: False]
    link [5297] 'Advanced Search'
    button [4600] 'Search' [disabled: True]

    tablist [4177]
        tabpanel
            menu "[4331] 'Beauty & Personal Care'; 
                  [4327] 'Sports & Outdoors'; 
                  [4323] 'Clothing, Shoes & Jewelry'; 
                  [4319] 'Home & Kitchen'; 
                  [4315] 'Office Products'; 
                  [4311] 'Tools & Home Improvement'; 
                  [4307] 'Health & Household'; 
                  [4303] 'Patio, Lawn & Garden'; 
                  [4299] 'Electronics'; 
                  [4295] 'Cell Phones & Accessories'; 
                  [4291] 'Video Games'; 
                  [4287] 'Grocery & Gourmet Food'"

    main
        heading 'My Wish List'
        LineBreak [4447]
        link [4571] 'My Account'
        link [4572] 'My Orders'
        link [4573] 'My Downloadable Products'
        link [4576] 'Address Book'
        link [4577] 'Account Information'
        link [4578] 'Stored Payment Methods'
        link [4580] 'My Product Reviews'
        link [4581] 'Newsletter Subscriptions'

        heading 'Compare Products'
        text 'You have no items to compare.'

        heading 'My Wish List'

    contentinfo
        textbox [4584] 'Sign Up for Our Newsletter:' [required: False]
        button [4543] 'Subscribe'
        link [4544] 'Privacy and Cookie Policy'
        link [4545] 'Search Terms'
        link [4546] 'Advanced Search'
        link [4547] 'Contact Us'
        text 'Copyright 2013-present Magento, Inc. All rights reserved.'
        text 'Help Us Keep Magento Healthy'
        link [4507] 'Report All Bugs'
\end{lstlisting}
\textbf{Action 1:} type [4597] [HONGJ Hawaiian Beach Outfits Set for Mens] [1]
\textbf{Reason 1:} 
\begin{lstlisting}
Using the search functionality could directly find the desired product if it exists.
\end{lstlisting}

\textbf{Observation 2:} 
\begin{lstlisting}
RootWebArea "Search results for: 'HONGJ Hawaiian Beach Outfits Set for Mens'"
    link [7082] 'My Account'
    link [7080] 'My Wish List'
    link [7084] 'Sign Out'
    text 'Welcome to One Stop Market'
    link [7017] 'Skip to Content'
    link [7026] 'store logo'
    link [7091] 'My Cart'
    combobox [7245] 'Search' [required: False]
        text 'HONGJ Hawaiian Beach Outfits Set for Mens'
    link [7956] 'Advanced Search'
    button [7248] 'Search' [disabled: True]

    tablist [6433]
        tabpanel
            menu "[6941] 'Beauty & Personal Care'; 
                  [6937] 'Sports & Outdoors'; 
                  [6933] 'Clothing, Shoes & Jewelry'; 
                  [6929] 'Home & Kitchen'; 
                  [6925] 'Office Products'; 
                  [6921] 'Tools & Home Improvement'; 
                  [6917] 'Health & Household'; 
                  [6913] 'Patio, Lawn & Garden'; 
                  [6909] 'Electronics'; 
                  [6905] 'Cell Phones & Accessories'; 
                  [6901] 'Video Games'; 
                  [6897] 'Grocery & Gourmet Food'"

    link [7049] 'Home'
    
    main
        heading "Search results for: 'HONGJ Hawaiian Beach Outfits Set for Mens'"
        strong [7102] 'Grid'
        link [7103] 'View as List'
        text 'Items'
        text '12'
        text 'of'
        text '38344'
        combobox "Sort By: [7204] 'Product Name' [selected: False]; 
                           [7205] 'Price' [selected: False]; 
                           [7206] 'Relevance' [selected: True]"
        link [6718] 'Set Ascending Direction'

        DescriptionList
            DescriptionListTerm 
                "Related search terms: 
                link [7211] 'set of 2 pillow cover 20x40'; 
                link [7212] 'set of 2 pillow cover silver grey 20x40'; 
                link [7213] 'set of 2 pillow cover blue grey 20x20'"

        link [7962] 'HONGJ Hawaiian Beach Outfits Set for Mens, Summer Tropical Tree Printed Relaxed-fit Hawaii Shirts Shorts 2 Piece Suits'
        text '18.98'
        button [8110] 'Add to Cart'
        button [6728] 'Add to Wish List'
        button [8111] 'Add to Compare'

        link [7972] 'HONGJ 2022 New Mens Cotton Linen Shirts, Spring Plus Size Long Sleeve V Neck Hawaiian Shirts Breathable Loose Beach Tops'
        text '5.98'
        button [8118] 'Add to Cart'
        button [6736] 'Add to Wish List'
        button [8119] 'Add to Compare'

        link [7982] 'HONGJ Hawaiian Shirts for Mens, Spring Long Sleeve Street Vintage Boho Printed Button Down Patchwork Casual Shirt'
        text '11.98'
        button [8126] 'Add to Cart'
        button [6744] 'Add to Wish List'
        button [8127] 'Add to Compare'

        link [7992] 'MOOKO Mens Summer Short Sleeve Hawaiian Shirt Sets 2 Pieces Flower Beach Outfits Button Down Shirt Shorts Suit Tracksuit'
        text '10.64'
        button [8134] 'Add to Cart'
        button [6752] 'Add to Wish List'
        button [8135] 'Add to Compare'

        link [8002] 'Men 2 Piece Outfit Summer Stylish Casual Hawaiian Shirts Graphic Tees Tank Tops Shorts Set Beach Tracksuit Pockets'
        LayoutTable
            text 'Rating: 62'
            link [8138] '12 Reviews'
        text '4.71'
        button [8144] 'Add to Cart'
        button [6766] 'Add to Wish List'
        button [8145] 'Add to Compare'

        list 'Page'
            text "You're currently reading page"
            text '1'
            link [7945] 'Page 2'
            link [7946] 'Page 3'
            link [7947] 'Page 4'
            link [7948] 'Page 5'
            link [6828] 'Page Next'

        combobox "Show: [7222] '12' [selected: True]; 
                           [7223] '24' [selected: False]; 
                           [7224] '36' [selected: False]"
        text 'per page'
        text 'Shop By'
        
        heading 'Shopping Options'
        DescriptionList
            heading 'Category'
            DescriptionListDetail
                link [6837] 'Beauty & Personal Care (8113 items)'
                link [6842] 'Sports & Outdoors (691 items)'
                link [6847] 'Clothing, Shoes & Jewelry (15836 items)'
                link [6852] 'Home & Kitchen (6007 items)'
                link [6857] 'Office Products (191 items)'
                link [6862] 'Tools & Home Improvement (807 items)'
                link [6867] 'Health & Household (115 items)'
                link [6872] 'Patio, Lawn & Garden (151 items)'
                link [6877] 'Electronics (3584 items)'
                link [6882] 'Cell Phones & Accessories (640 items)'
                link [6887] 'Video Games (231 items)'
                link [6892] 'Grocery & Gourmet Food (1978 items)'

        heading 'Compare Products'
        text 'You have no items to compare.'
        heading 'My Wish List'
        text 'You have no items in your wish list.'

    contentinfo
        textbox [7232] 'Sign Up for Our Newsletter:' [required: False]
        button [7174] 'Subscribe'
        link [7175] 'Privacy and Cookie Policy'
        link [7176] 'Search Terms'
        link [7177] 'Advanced Search'
        link [7178] 'Contact Us'
        text 'Copyright 2013-present Magento, Inc. All rights reserved.'
        text 'Help Us Keep Magento Healthy'
        link [7079] 'Report All Bugs'
\end{lstlisting}
\textbf{Action 2:} click [6728]
\textbf{Reason 2:} 
\begin{lstlisting}
We have successfully found the HONGJ Hawaiian Beach Outfits Set for Mens that we were looking for. The next step is to add this item to the wish list as per the task objective. The \"Add to Wish List\" button [6728] is available next to the item.
\end{lstlisting}

\textbf{Observation 3:} 
\begin{lstlisting}
RootWebArea "My Wish List"
    link [10068] 'My Account'
    link [10064] 'My Wish List 1 item'
    link [10070] 'Sign Out'
    text 'Welcome to One Stop Market'
    link [9948] 'Skip to Content'
    link [9957] 'store logo'
    link [10077] 'My Cart'
    combobox [10199] 'Search' [required: False]
    link [10938] 'Advanced Search'
    button [10202] 'Search' [disabled: True]

    tablist [9700]
        tabpanel
            menu "[9877] 'Beauty & Personal Care'; 
                  [9873] 'Sports & Outdoors'; 
                  [9869] 'Clothing, Shoes & Jewelry'; 
                  [9865] 'Home & Kitchen'; 
                  [9861] 'Office Products'; 
                  [9857] 'Tools & Home Improvement'; 
                  [9853] 'Health & Household'; 
                  [9849] 'Patio, Lawn & Garden'; 
                  [9845] 'Electronics'; 
                  [9841] 'Cell Phones & Accessories'; 
                  [9837] 'Video Games'; 
                  [9833] 'Grocery & Gourmet Food'"

    main
        alert [live: assertive] [atomic: True] [relevant: additions text]
            text "HONGJ Hawaiian Beach Outfits Set for Mens, Summer Tropical Tree Printed Relaxed-fit Hawaii Shirts Shorts 2 Piece Suits has been added to your Wish List. Click"
            link [10155] 'here'
            text 'to continue shopping.'

        heading 'My Wish List'
        combobox "[10158] '10' [selected: True]; 
                  [10159] '20' [selected: False]; 
                  [10160] '50' [selected: False]"

        link [10892] 'HONGJ Hawaiian Beach Outfits Set for Mens, Summer Tropical Tree Printed Relaxed-fit Hawaii Shirts Shorts 2 Piece Suits'
        textbox [10903] 'Comment' [required: False]

        group [10906]
            spinbutton [10946] 'Qty' [required: False] [valuemin: 0] [valuemax: 0] [valuetext: ] 
                text '1'
            button [10965] 'Add to Cart'

        link [10909] 'Edit'
        link [10910] 'Remove item'
        button [10094] 'Update Wish List'
        button [10095] 'Share Wish List'
        button [10096] 'Add All to Cart'

        LineBreak [10001]
        combobox "[10169] '10' [selected: True]; 
                  [10170] '20' [selected: False]; 
                  [10171] '50' [selected: False]"

        link [10172] 'My Account'
        link [10173] 'My Orders'
        link [10174] 'My Downloadable Products'
        link [10177] 'Address Book'
        link [10178] 'Account Information'
        link [10179] 'Stored Payment Methods'
        link [10181] 'My Product Reviews'
        link [10182] 'Newsletter Subscriptions'

        heading 'Compare Products'
        text 'You have no items to compare.'

        heading 'My Wish List'
        text '1 item'
        link [10925] 'HONGJ Hawaiian Beach Outfits Set for Mens, Summer Tropical Tree Printed Relaxed-fit Hawaii Shirts Shorts 2 Piece Suits'
        link [10956] 'Add to Cart'
        link [9830] 'Remove This Item'
        link [10042] 'Go to Wish List'

    contentinfo
        textbox [10186] 'Sign Up for Our Newsletter:' [required: False]
        button [10127] 'Subscribe'
        link [10128] 'Privacy and Cookie Policy'
        link [10129] 'Search Terms'
        link [10130] 'Advanced Search'
        link [10131] 'Contact Us'
        text 'Copyright 2013-present Magento, Inc. All rights reserved.'
        text 'Help Us Keep Magento Healthy'
        link [10063] 'Report All Bugs'
\end{lstlisting}
\textbf{Action 3:} stop [The HONGJ Hawaiian Beach Outfits Set for Mens has been successfully added to the Wish List. The task is complete.]
\textbf{Reason 3:} 
\begin{lstlisting}
The objective has been successfully completed. The HONGJ Hawaiian Beach Outfits Set has been added to the Wish List, as confirmed by the alert message and the updated page content.
\end{lstlisting}
\textbf{success:}
1.0

\textbf{num actions:}
4
\end{tcolorbox}

%% file: custom.bbl
\begin{thebibliography}{89}
\providecommand{\natexlab}[1]{#1}

\bibitem[{Anand et~al.(2023)Anand, Nussbaum, Duderstadt, Schmidt, and Mulyar}]{anand2023gpt4all}
Yuvanesh Anand, Zach Nussbaum, Brandon Duderstadt, Benjamin Schmidt, and Andriy Mulyar. 2023.
\newblock \href {https://github.com/nomic-ai/gpt4all} {Gpt4all: Training an assistant-style chatbot with large scale data distillation from gpt-3.5-turbo}.
\newblock Technical report.

\bibitem[{Antoniades et~al.(2024)Antoniades, {\"O}rwall, Zhang, Xie, Goyal, and Wang}]{antoniades2024swe}
Antonis Antoniades, Albert {\"O}rwall, Kexun Zhang, Yuxi Xie, Anirudh Goyal, and William Wang. 2024.
\newblock Swe-search: Enhancing software agents with monte carlo tree search and iterative refinement.
\newblock \emph{arXiv preprint arXiv:2410.20285}.

\bibitem[{Caccia et~al.()Caccia, Thakkar, Boisvert, De~Chezelles, Pich{\'e}, Chapados, Drouin, Gasse, and Lacoste}]{caccia2024fine}
Massimo Caccia, Megh Thakkar, L{\'e}o Boisvert, Thibault Le~Sellier De~Chezelles, Alexandre Pich{\'e}, Nicolas Chapados, Alexandre Drouin, Maxime Gasse, and Alexandre Lacoste.
\newblock Fine-tuning web agents: It works, but it's trickier than you think.
\newblock In \emph{NeurIPS 2024 Workshop on Open-World Agents}.

\bibitem[{Cao et~al.(2024)Cao, Lei, Wu, Chen, Fu, Gao, Xiong, Zhang, Mao, Hu et~al.}]{cao2024spider2}
Ruisheng Cao, Fangyu Lei, Haoyuan Wu, Jixuan Chen, Yeqiao Fu, Hongcheng Gao, Xinzhuang Xiong, Hanchong Zhang, Yuchen Mao, Wenjing Hu, et~al. 2024.
\newblock Spider2-v: How far are multimodal agents from automating data science and engineering workflows?
\newblock \emph{arXiv preprint arXiv:2407.10956}.

\bibitem[{Chae et~al.(2025)Chae, Kim, iunn Ong, Gwak, Song, Kim, Kim, Lee, and Yeo}]{chae2025web}
Hyungjoo Chae, Namyoung Kim, Kai~Tzu iunn Ong, Minju Gwak, Gwanwoo Song, Jihoon Kim, Sunghwan Kim, Dongha Lee, and Jinyoung Yeo. 2025.
\newblock \href {https://openreview.net/forum?id=moWiYJuSGF} {Web agents with world models: Learning and leveraging environment dynamics in web navigation}.
\newblock In \emph{The Thirteenth International Conference on Learning Representations}.

\bibitem[{Chen et~al.(2023)Chen, Shu, Shareghi, Collier, Narasimhan, and Yao}]{chen2023fireact}
Baian Chen, Chang Shu, Ehsan Shareghi, Nigel Collier, Karthik Narasimhan, and Shunyu Yao. 2023.
\newblock Fireact: Toward language agent fine-tuning.
\newblock \emph{arXiv preprint arXiv:2310.05915}.

\bibitem[{Chen et~al.(2017)Chen, Choi, Yu, Han, and Chandraker}]{10.5555/3294771.3294842}
Guobin Chen, Wongun Choi, Xiang Yu, Tony Han, and Manmohan Chandraker. 2017.
\newblock Learning efficient object detection models with knowledge distillation.
\newblock In \emph{Proceedings of the 31st International Conference on Neural Information Processing Systems}, NIPS'17, page 742–751, Red Hook, NY, USA. Curran Associates Inc.

\bibitem[{Chen et~al.(2025)Chen, You, Li, yitong guan, Qian, Zhao, Yang, Xie, Liu, and Sun}]{chen2025internet}
Weize Chen, Ziming You, Ran Li, yitong guan, Chen Qian, Chenyang Zhao, Cheng Yang, Ruobing Xie, Zhiyuan Liu, and Maosong Sun. 2025.
\newblock \href {https://openreview.net/forum?id=o1Et3MogPw} {Internet of agents: Weaving a web of heterogeneous agents for collaborative intelligence}.
\newblock In \emph{The Thirteenth International Conference on Learning Representations}.

\bibitem[{Chezelles et~al.(2024)Chezelles, Gasse, Drouin, Caccia, Boisvert, Thakkar, Marty, Assouel, Shayegan, Jang, Lù, Yoran, Kong, Xu, Reddy, Cappart, Neubig, Salakhutdinov, Chapados, and Lacoste}]{dechezelles2024browsergym}
Thibault Le Sellier~De Chezelles, Maxime Gasse, Alexandre Drouin, Massimo Caccia, Léo Boisvert, Megh Thakkar, Tom Marty, Rim Assouel, Sahar~Omidi Shayegan, Lawrence~Keunho Jang, Xing~Han Lù, Ori Yoran, Dehan Kong, Frank~F. Xu, Siva Reddy, Quentin Cappart, Graham Neubig, Ruslan Salakhutdinov, Nicolas Chapados, and Alexandre Lacoste. 2024.
\newblock \href {https://arxiv.org/abs/2412.05467} {The browsergym ecosystem for web agent research}.
\newblock \emph{Preprint}, arXiv:2412.05467.

\bibitem[{Deng et~al.(2023)Deng, Gu, Zheng, Chen, Stevens, Wang, Sun, and Su}]{deng2023mind2web}
Xiang Deng, Yu~Gu, Boyuan Zheng, Shijie Chen, Samuel Stevens, Boshi Wang, Huan Sun, and Yu~Su. 2023.
\newblock \href {https://arxiv.org/abs/2306.06070} {Mind2web: Towards a generalist agent for the web}.
\newblock \emph{Preprint}, arXiv:2306.06070.

\bibitem[{Du et~al.(2023)Du, Li, Torralba, Tenenbaum, and Mordatch}]{du2023improving}
Yilun Du, Shuang Li, Antonio Torralba, Joshua~B Tenenbaum, and Igor Mordatch. 2023.
\newblock Improving factuality and reasoning in language models through multiagent debate.
\newblock \emph{arXiv preprint arXiv:2305.14325}.

\bibitem[{Fourney et~al.(2024)Fourney, Bansal, Mozannar, Tan, Salinas, Erkang, Zhu, Niedtner, Proebsting, Bassman, Gerrits, Alber, Chang, Loynd, West, Dibia, Awadallah, Kamar, Hosn, and Amershi}]{fourney2024magenticone}
Adam Fourney, Gagan Bansal, Hussein Mozannar, Cheng Tan, Eduardo Salinas, Erkang, Zhu, Friederike Niedtner, Grace Proebsting, Griffin Bassman, Jack Gerrits, Jacob Alber, Peter Chang, Ricky Loynd, Robert West, Victor Dibia, Ahmed Awadallah, Ece Kamar, Rafah Hosn, and Saleema Amershi. 2024.
\newblock \href {https://arxiv.org/abs/2411.04468} {Magentic-one: A generalist multi-agent system for solving complex tasks}.
\newblock \emph{Preprint}, arXiv:2411.04468.

\bibitem[{Fu et~al.(2025)Fu, He, Wang, Hong, Gongque, Zeng, Wang, Wang, Cai, and Xu}]{fu2025agentrefine}
Dayuan Fu, Keqing He, Yejie Wang, Wentao Hong, Zhuoma Gongque, Weihao Zeng, Wei Wang, Jingang Wang, Xunliang Cai, and Weiran Xu. 2025.
\newblock Agentrefine: Enhancing agent generalization through refinement tuning.
\newblock \emph{arXiv preprint arXiv:2501.01702}.

\bibitem[{Fu et~al.(2023)Fu, Peng, Ou, Sabharwal, and Khot}]{fu2023specializing}
Yao Fu, Hao Peng, Litu Ou, Ashish Sabharwal, and Tushar Khot. 2023.
\newblock \href {https://arxiv.org/abs/2301.12726} {Specializing smaller language models towards multi-step reasoning}.
\newblock \emph{Preprint}, arXiv:2301.12726.

\bibitem[{Gonzalez-Pumariega et~al.(2025)Gonzalez-Pumariega, Yean, Sunkara, and Choudhury}]{gonzalez-pumariega2025robotouille}
Gonzalo Gonzalez-Pumariega, Leong~Su Yean, Neha Sunkara, and Sanjiban Choudhury. 2025.
\newblock \href {https://openreview.net/forum?id=OhUoTMxFIH} {Robotouille: An asynchronous planning benchmark for {LLM} agents}.
\newblock In \emph{The Thirteenth International Conference on Learning Representations}.

\bibitem[{Gou et~al.(2021)Gou, Yu, Maybank, and Tao}]{Gou_2021}
Jianping Gou, Baosheng Yu, Stephen~J. Maybank, and Dacheng Tao. 2021.
\newblock \href {https://doi.org/10.1007/s11263-021-01453-z} {Knowledge distillation: A survey}.
\newblock \emph{International Journal of Computer Vision}, 129(6):1789–1819.

\bibitem[{Gu et~al.(2024)Gu, Jiang, Shi, Tan, Zhai, Xu, Li, Shen, Ma, Liu et~al.}]{gu2024survey}
Jiawei Gu, Xuhui Jiang, Zhichao Shi, Hexiang Tan, Xuehao Zhai, Chengjin Xu, Wei Li, Yinghan Shen, Shengjie Ma, Honghao Liu, et~al. 2024.
\newblock A survey on llm-as-a-judge.
\newblock \emph{arXiv preprint arXiv:2411.15594}.

\bibitem[{Guo et~al.(2025)Guo, Yang, Zhang, Song, Zhang, Xu, Zhu, Ma, Wang, Bi et~al.}]{guo2025deepseek}
Daya Guo, Dejian Yang, Haowei Zhang, Junxiao Song, Ruoyu Zhang, Runxin Xu, Qihao Zhu, Shirong Ma, Peiyi Wang, Xiao Bi, et~al. 2025.
\newblock Deepseek-r1: Incentivizing reasoning capability in llms via reinforcement learning.
\newblock \emph{arXiv preprint arXiv:2501.12948}.

\bibitem[{Gur et~al.(2023)Gur, Furuta, Huang, Safdari, Matsuo, Eck, and Faust}]{gur2023real}
Izzeddin Gur, Hiroki Furuta, Austin Huang, Mustafa Safdari, Yutaka Matsuo, Douglas Eck, and Aleksandra Faust. 2023.
\newblock A real-world webagent with planning, long context understanding, and program synthesis.
\newblock \emph{arXiv preprint arXiv:2307.12856}.

\bibitem[{Hinton et~al.(2015)Hinton, Vinyals, and Dean}]{hinton2015distilling}
Geoffrey Hinton, Oriol Vinyals, and Jeff Dean. 2015.
\newblock \href {https://arxiv.org/abs/1503.02531} {Distilling the knowledge in a neural network}.
\newblock \emph{Preprint}, arXiv:1503.02531.

\bibitem[{Hong et~al.(2024)Hong, Wang, Lv, Xu, Yu, Ji, Wang, Wang, Dong, Ding et~al.}]{hong2024cogagent}
Wenyi Hong, Weihan Wang, Qingsong Lv, Jiazheng Xu, Wenmeng Yu, Junhui Ji, Yan Wang, Zihan Wang, Yuxiao Dong, Ming Ding, et~al. 2024.
\newblock Cogagent: A visual language model for gui agents.
\newblock In \emph{Proceedings of the IEEE/CVF Conference on Computer Vision and Pattern Recognition}, pages 14281--14290.

\bibitem[{Hou et~al.(2024)Hou, Yang, Jiao, Wang, Tu, and Zhao}]{hou2024coact}
Xinming Hou, Mingming Yang, Wenxiang Jiao, Xing Wang, Zhaopeng Tu, and Wayne~Xin Zhao. 2024.
\newblock \href {https://arxiv.org/abs/2406.13381} {Coact: A global-local hierarchy for autonomous agent collaboration}.
\newblock \emph{Preprint}, arXiv:2406.13381.

\bibitem[{Hsieh et~al.(2023)Hsieh, Li, Yeh, Nakhost, Fujii, Ratner, Krishna, Lee, and Pfister}]{hsieh2023distilling}
Cheng-Yu Hsieh, Chun-Liang Li, Chih-Kuan Yeh, Hootan Nakhost, Yasuhisa Fujii, Alexander Ratner, Ranjay Krishna, Chen-Yu Lee, and Tomas Pfister. 2023.
\newblock \href {https://arxiv.org/abs/2305.02301} {Distilling step-by-step! outperforming larger language models with less training data and smaller model sizes}.
\newblock \emph{Preprint}, arXiv:2305.02301.

\bibitem[{Hu et~al.(2024)Hu, Zhao, Xu, Sun, Lou, Lin, Luo, and Rajmohan}]{hu2024agentgen}
Mengkang Hu, Pu~Zhao, Can Xu, Qingfeng Sun, Jianguang Lou, Qingwei Lin, Ping Luo, and Saravan Rajmohan. 2024.
\newblock \href {https://arxiv.org/abs/2408.00764} {Agentgen: Enhancing planning abilities for large language model based agent via environment and task generation}.
\newblock \emph{Preprint}, arXiv:2408.00764.

\bibitem[{Huang et~al.(2024{\natexlab{a}})Huang, Gu, Peng, Liang, Li, Xiao, Wen, and Chen}]{DBLP:conf/emnlp/HuangGPLLXWC24}
Wenhao Huang, Zhouhong Gu, Chenghao Peng, Jiaqing Liang, Zhixu Li, Yanghua Xiao, Liqian Wen, and Zulong Chen. 2024{\natexlab{a}}.
\newblock \href {https://aclanthology.org/2024.emnlp-main.141} {Autoscraper: A progressive understanding web agent for web scraper generation}.
\newblock In \emph{EMNLP}, pages 2371--2389.

\bibitem[{Huang et~al.(2024{\natexlab{b}})Huang, Peng, Li, Liang, Xiao, Wen, and Chen}]{DBLP:journals/corr/abs-2404-12753}
Wenhao Huang, Chenghao Peng, Zhixu Li, Jiaqing Liang, Yanghua Xiao, Liqian Wen, and Zulong Chen. 2024{\natexlab{b}}.
\newblock \href {https://doi.org/10.48550/arXiv.2404.12753} {Autocrawler: A progressive understanding web agent for web crawler generation}.
\newblock \emph{CoRR}, abs/2404.12753.

\bibitem[{Huang et~al.(2022)Huang, Chen, Yu, and McKeown}]{huang2022incontext}
Yukun Huang, Yanda Chen, Zhou Yu, and Kathleen McKeown. 2022.
\newblock \href {https://arxiv.org/abs/2212.10670} {In-context learning distillation: Transferring few-shot learning ability of pre-trained language models}.
\newblock \emph{Preprint}, arXiv:2212.10670.

\bibitem[{Jang et~al.(2025)Jang, Li, Zhao, Ding, Lin, Liang, Bonatti, and Koishida}]{jang2025videowebarena}
Lawrence~Keunho Jang, Yinheng Li, Dan Zhao, Charles Ding, Justin Lin, Paul~Pu Liang, Rogerio Bonatti, and Kazuhito Koishida. 2025.
\newblock \href {https://openreview.net/forum?id=unDQOUah0F} {Videowebarena: Evaluating long context multimodal agents with video understanding web tasks}.
\newblock In \emph{The Thirteenth International Conference on Learning Representations}.

\bibitem[{Jimenez et~al.(2023)Jimenez, Yang, Wettig, Yao, Pei, Press, and Narasimhan}]{jimenez2023swe}
Carlos~E Jimenez, John Yang, Alexander Wettig, Shunyu Yao, Kexin Pei, Ofir Press, and Karthik Narasimhan. 2023.
\newblock Swe-bench: Can language models resolve real-world github issues?
\newblock \emph{arXiv preprint arXiv:2310.06770}.

\bibitem[{Kim et~al.(2020)Kim, Park, and Kwak}]{kim2020paraphrasing}
Jangho Kim, SeongUk Park, and Nojun Kwak. 2020.
\newblock \href {https://arxiv.org/abs/1802.04977} {Paraphrasing complex network: Network compression via factor transfer}.
\newblock \emph{Preprint}, arXiv:1802.04977.

\bibitem[{Koh et~al.(2024)Koh, McAleer, Fried, and Salakhutdinov}]{koh2024tree}
Jing~Yu Koh, Stephen McAleer, Daniel Fried, and Ruslan Salakhutdinov. 2024.
\newblock Tree search for language model agents.
\newblock \emph{arXiv preprint arXiv:2407.01476}.

\bibitem[{Kumar et~al.(2025)Kumar, Lau, Vijayakumar, Trinh, Chang, Robinson, Zhou, Fredrikson, Hendryx, Yue, and Wang}]{kumar2025aligned}
Priyanshu Kumar, Elaine Lau, Saranya Vijayakumar, Tu~Trinh, Elaine~T Chang, Vaughn Robinson, Shuyan Zhou, Matt Fredrikson, Sean~M. Hendryx, Summer Yue, and Zifan Wang. 2025.
\newblock \href {https://openreview.net/forum?id=NsFZZU9gvk} {Aligned {LLM}s are not aligned browser agents}.
\newblock In \emph{The Thirteenth International Conference on Learning Representations}.

\bibitem[{Lai et~al.(2024)Lai, Liu, Iong, Yao, Chen, Shen, Yu, Zhang, Zhang, Dong et~al.}]{lai2024autowebglm}
Hanyu Lai, Xiao Liu, Iat~Long Iong, Shuntian Yao, Yuxuan Chen, Pengbo Shen, Hao Yu, Hanchen Zhang, Xiaohan Zhang, Yuxiao Dong, et~al. 2024.
\newblock Autowebglm: A large language model-based web navigating agent.
\newblock In \emph{Proceedings of the 30th ACM SIGKDD Conference on Knowledge Discovery and Data Mining}, pages 5295--5306.

\bibitem[{Lee et~al.(2025)Lee, Lee, Kim, Tack, Shin, Teh, and Lee}]{lee2025learning}
Dongjun Lee, Juyong Lee, Kyuyoung Kim, Jihoon Tack, Jinwoo Shin, Yee~Whye Teh, and Kimin Lee. 2025.
\newblock \href {https://openreview.net/forum?id=3Gzz7ZQLiz} {Learning to contextualize web pages for enhanced decision making by {LLM} agents}.
\newblock In \emph{The Thirteenth International Conference on Learning Representations}.

\bibitem[{Levy et~al.(2024)Levy, Wiesel, Marreed, Oved, Yaeli, and Shlomov}]{levy2024stwebagentbench}
Ido Levy, Ben Wiesel, Sami Marreed, Alon Oved, Avi Yaeli, and Segev Shlomov. 2024.
\newblock \href {https://arxiv.org/abs/2410.06703} {St-webagentbench: A benchmark for evaluating safety and trustworthiness in web agents}.
\newblock \emph{Preprint}, arXiv:2410.06703.

\bibitem[{Li et~al.(2025{\natexlab{a}})Li, Xie, Li, Tsung, Ding, and Li}]{li2025agentoriented}
Ao~Li, Yuexiang Xie, Songze Li, Fugee Tsung, Bolin Ding, and Yaliang Li. 2025{\natexlab{a}}.
\newblock \href {https://openreview.net/forum?id=EqcLAU6gyU} {Agent-oriented planning in multi-agent systems}.
\newblock In \emph{The Thirteenth International Conference on Learning Representations}.

\bibitem[{Li et~al.(2024{\natexlab{a}})Li, Chen, Li, Wang, Li, Chen, and Zhang}]{li2024mixed}
Chenglin Li, Qianglong Chen, Liangyue Li, Caiyu Wang, Yicheng Li, Zulong Chen, and Yin Zhang. 2024{\natexlab{a}}.
\newblock \href {https://arxiv.org/abs/2312.10730} {Mixed distillation helps smaller language model better reasoning}.
\newblock \emph{Preprint}, arXiv:2312.10730.

\bibitem[{Li et~al.(2024{\natexlab{b}})Li, Jiang, Huang, Beigi, Zhao, Tan, Bhattacharjee, Jiang, Chen, Wu et~al.}]{li2024generation}
Dawei Li, Bohan Jiang, Liangjie Huang, Alimohammad Beigi, Chengshuai Zhao, Zhen Tan, Amrita Bhattacharjee, Yuxuan Jiang, Canyu Chen, Tianhao Wu, et~al. 2024{\natexlab{b}}.
\newblock From generation to judgment: Opportunities and challenges of llm-as-a-judge.
\newblock \emph{arXiv preprint arXiv:2411.16594}.

\bibitem[{Li et~al.(2020)Li, He, Zhou, Zhang, and Baldridge}]{li2020mapping}
Yang Li, Jiacong He, Xin Zhou, Yuan Zhang, and Jason Baldridge. 2020.
\newblock Mapping natural language instructions to mobile ui action sequences.
\newblock \emph{arXiv preprint arXiv:2005.03776}.

\bibitem[{Li et~al.(2025{\natexlab{b}})Li, Li, Wang, Jiang, Zhang, Zheng, Wang, Zheng, Huang, Zhou, and Yu}]{li2025benchmarking}
Yangning Li, Yinghui Li, Xinyu Wang, Yong Jiang, Zhen Zhang, Xinran Zheng, Hui Wang, Hai-Tao Zheng, Fei Huang, Jingren Zhou, and Philip~S. Yu. 2025{\natexlab{b}}.
\newblock \href {https://openreview.net/forum?id=VvDEuyVXkG} {Benchmarking multimodal retrieval augmented generation with dynamic {VQA} dataset and self-adaptive planning agent}.
\newblock In \emph{The Thirteenth International Conference on Learning Representations}.

\bibitem[{Liang et~al.(2024)Liang, He, Jiao, Wang, Wang, Wang, Yang, Shi, and Tu}]{liang-etal-2024-encouraging}
Tian Liang, Zhiwei He, Wenxiang Jiao, Xing Wang, Yan Wang, Rui Wang, Yujiu Yang, Shuming Shi, and Zhaopeng Tu. 2024.
\newblock \href {https://doi.org/10.18653/v1/2024.emnlp-main.992} {Encouraging divergent thinking in large language models through multi-agent debate}.
\newblock In \emph{Proceedings of the 2024 Conference on Empirical Methods in Natural Language Processing}, pages 17889--17904, Miami, Florida, USA. Association for Computational Linguistics.

\bibitem[{Light et~al.(2025)Light, Cai, Chen, Wang, Chen, Cheng, Yue, and Hu}]{light2025strategist}
Jonathan Light, Min Cai, Weiqin Chen, Guanzhi Wang, Xiusi Chen, Wei Cheng, Yisong Yue, and Ziniu Hu. 2025.
\newblock \href {https://openreview.net/forum?id=gfI9v7AbFg} {Strategist: Self-improvement of {LLM} decision making via bi-level tree search}.
\newblock In \emph{The Thirteenth International Conference on Learning Representations}.

\bibitem[{Liu et~al.(2025)Liu, Ou, Song, Qu, Lam, Xiong, Chen, Neubig, and Yue}]{liu2025harnessing}
Junpeng Liu, Tianyue Ou, Yifan Song, Yuxiao Qu, Wai Lam, Chenyan Xiong, Wenhu Chen, Graham Neubig, and Xiang Yue. 2025.
\newblock \href {https://openreview.net/forum?id=IIsTO4P3Ag} {Harnessing webpage {UI}s for text-rich visual understanding}.
\newblock In \emph{The Thirteenth International Conference on Learning Representations}.

\bibitem[{Magister et~al.(2023)Magister, Mallinson, Adamek, Malmi, and Severyn}]{magister2023teaching}
Lucie~Charlotte Magister, Jonathan Mallinson, Jakub Adamek, Eric Malmi, and Aliaksei Severyn. 2023.
\newblock \href {https://arxiv.org/abs/2212.08410} {Teaching small language models to reason}.
\newblock \emph{Preprint}, arXiv:2212.08410.

\bibitem[{Mirzadeh et~al.(2019)Mirzadeh, Farajtabar, Li, Levine, Matsukawa, and Ghasemzadeh}]{mirzadeh2019improved}
Seyed-Iman Mirzadeh, Mehrdad Farajtabar, Ang Li, Nir Levine, Akihiro Matsukawa, and Hassan Ghasemzadeh. 2019.
\newblock \href {https://arxiv.org/abs/1902.03393} {Improved knowledge distillation via teacher assistant}.
\newblock \emph{Preprint}, arXiv:1902.03393.

\bibitem[{Mukherjee et~al.(2023)Mukherjee, Mitra, Jawahar, Agarwal, Palangi, and Awadallah}]{mukherjee2023orca}
Subhabrata Mukherjee, Arindam Mitra, Ganesh Jawahar, Sahaj Agarwal, Hamid Palangi, and Ahmed Awadallah. 2023.
\newblock \href {https://arxiv.org/abs/2306.02707} {Orca: Progressive learning from complex explanation traces of gpt-4}.
\newblock \emph{Preprint}, arXiv:2306.02707.

\bibitem[{Pan et~al.(2024{\natexlab{a}})Pan, Zhang, Tomlin, Zhou, Levine, and Suhr}]{pan2024autonomous}
Jiayi Pan, Yichi Zhang, Nicholas Tomlin, Yifei Zhou, Sergey Levine, and Alane Suhr. 2024{\natexlab{a}}.
\newblock Autonomous evaluation and refinement of digital agents.
\newblock \emph{arXiv preprint arXiv:2404.06474}.

\bibitem[{Pan et~al.(2024{\natexlab{b}})Pan, Kong, Zhou, Cui, Leng, Jiang, Liu, Shang, Zhou, Wu, and Wu}]{pan2024webcanvas}
Yichen Pan, Dehan Kong, Sida Zhou, Cheng Cui, Yifei Leng, Bing Jiang, Hangyu Liu, Yanyi Shang, Shuyan Zhou, Tongshuang Wu, and Zhengyang Wu. 2024{\natexlab{b}}.
\newblock \href {https://arxiv.org/abs/2406.12373} {Webcanvas: Benchmarking web agents in online environments}.
\newblock \emph{Preprint}, arXiv:2406.12373.

\bibitem[{Patel et~al.(2025)Patel, Hofmarcher, Leoveanu-Condrei, Dinu, Callison-Burch, and Hochreiter}]{patel2025large}
Ajay Patel, Markus Hofmarcher, Claudiu Leoveanu-Condrei, Marius-Constantin Dinu, Chris Callison-Burch, and Sepp Hochreiter. 2025.
\newblock \href {https://openreview.net/forum?id=jwME4SY0an} {Large language models can self-improve at web agent tasks}.

\bibitem[{Putta et~al.(2024)Putta, Mills, Garg, Motwani, Finn, Garg, and Rafailov}]{putta2024agentq}
Pranav Putta, Edmund Mills, Naman Garg, Sumeet Motwani, Chelsea Finn, Divyansh Garg, and Rafael Rafailov. 2024.
\newblock \href {https://arxiv.org/abs/2408.07199} {Agent q: Advanced reasoning and learning for autonomous ai agents}.
\newblock \emph{Preprint}, arXiv:2408.07199.

\bibitem[{Qi et~al.(2025)Qi, Liu, Iong, Lai, Sun, Sun, Yang, Yang, Yao, Xu, Tang, and Dong}]{qi2025webrl}
Zehan Qi, Xiao Liu, Iat~Long Iong, Hanyu Lai, Xueqiao Sun, Jiadai Sun, Xinyue Yang, Yu~Yang, Shuntian Yao, Wei Xu, Jie Tang, and Yuxiao Dong. 2025.
\newblock \href {https://openreview.net/forum?id=oVKEAFjEqv} {Web{RL}: Training {LLM} web agents via self-evolving online curriculum reinforcement learning}.
\newblock In \emph{The Thirteenth International Conference on Learning Representations}.

\bibitem[{Qi et~al.(2024)Qi, Liu, Iong, Lai, Sun, Zhao, Yang, Yang, Sun, Yao, Zhang, Xu, Tang, and Dong}]{qi2024webrl}
Zehan Qi, Xiao Liu, Iat~Long Iong, Hanyu Lai, Xueqiao Sun, Wenyi Zhao, Yu~Yang, Xinyue Yang, Jiadai Sun, Shuntian Yao, Tianjie Zhang, Wei Xu, Jie Tang, and Yuxiao Dong. 2024.
\newblock \href {https://arxiv.org/abs/2411.02337} {Webrl: Training llm web agents via self-evolving online curriculum reinforcement learning}.
\newblock \emph{Preprint}, arXiv:2411.02337.

\bibitem[{Sanh(2019)}]{sanh2019distilbert}
V~Sanh. 2019.
\newblock Distilbert, a distilled version of bert: smaller, faster, cheaper and lighter.
\newblock \emph{arXiv preprint arXiv:1910.01108}.

\bibitem[{SHEN et~al.(2025)SHEN, Li, Meng, Cai, Qi, Zhang, Xu, and Ma}]{shen2025shortcutsbench}
Haiyang SHEN, Yue Li, Desong Meng, Dongqi Cai, Sheng Qi, Li~Zhang, Mengwei Xu, and Yun Ma. 2025.
\newblock \href {https://openreview.net/forum?id=kKILfPkhSz} {Shortcutsbench: A large-scale real-world benchmark for {API}-based agents}.
\newblock In \emph{The Thirteenth International Conference on Learning Representations}.

\bibitem[{Shen et~al.(2024)Shen, Jain, Xiao, Amlekar, Hadji, Podolny, and Talwalkar}]{shen2024scribeagent}
Junhong Shen, Atishay Jain, Zedian Xiao, Ishan Amlekar, Mouad Hadji, Aaron Podolny, and Ameet Talwalkar. 2024.
\newblock \href {https://arxiv.org/abs/2411.15004} {Scribeagent: Towards specialized web agents using production-scale workflow data}.
\newblock \emph{Preprint}, arXiv:2411.15004.

\bibitem[{Shinn et~al.(2023)Shinn, Cassano, Labash, Gopinath, Narasimhan, and Yao}]{shinn2023reflexion}
Noah Shinn, Federico Cassano, Beck Labash, Ashwin Gopinath, Karthik Narasimhan, and Shunyu Yao. 2023.
\newblock Reflexion: Language agents with verbal reinforcement learning.(2023).
\newblock \emph{arXiv preprint cs.AI/2303.11366}.

\bibitem[{Sodhi et~al.(2024)Sodhi, Branavan, Artzi, and McDonald}]{sodhi2024step}
Paloma Sodhi, SRK Branavan, Yoav Artzi, and Ryan McDonald. 2024.
\newblock Step: Stacked llm policies for web actions.
\newblock In \emph{First Conference on Language Modeling}.

\bibitem[{Song et~al.(2024)Song, Xu, Zhou, and Neubig}]{song2024beyond}
Yueqi Song, Frank~F Xu, Shuyan Zhou, and Graham Neubig. 2024.
\newblock Beyond browsing: Api-based web agents.

\bibitem[{Sridhar et~al.(2025)Sridhar, Dutta, Jayaraman, and Lee}]{sridhar2025regent}
Kaustubh Sridhar, Souradeep Dutta, Dinesh Jayaraman, and Insup Lee. 2025.
\newblock \href {https://openreview.net/forum?id=NxyfSW6mLK} {{REGENT}: A retrieval-augmented generalist agent that can act in-context in new environments}.
\newblock In \emph{The Thirteenth International Conference on Learning Representations}.

\bibitem[{Su et~al.(2025)Su, Sun, Yoon, Yin, Yu, and Ar\i~k}]{su2025learn}
Hongjin Su, Ruoxi Sun, Jinsung Yoon, Pengcheng Yin, Tao Yu, and Sercan~\"O Ar\i~k. 2025.
\newblock Learn-by-interact: A data-centric framework for self-adaptive agents in realistic environments.
\newblock \emph{arXiv preprint arXiv:2501.10893}.

\bibitem[{Tan et~al.(2024)Tan, Li, Wang, Beigi, Jiang, Bhattacharjee, Karami, Li, Cheng, and Liu}]{tan2024large}
Zhen Tan, Dawei Li, Song Wang, Alimohammad Beigi, Bohan Jiang, Amrita Bhattacharjee, Mansooreh Karami, Jundong Li, Lu~Cheng, and Huan Liu. 2024.
\newblock Large language models for data annotation and synthesis: A survey.
\newblock In \emph{Proceedings of the 2024 Conference on Empirical Methods in Natural Language Processing}, pages 930--957.

\bibitem[{Tennant et~al.(2025)Tennant, Hailes, and Musolesi}]{tennant2025moral}
Elizaveta Tennant, Stephen Hailes, and Mirco Musolesi. 2025.
\newblock \href {https://openreview.net/forum?id=MeGDmZjUXy} {Moral alignment for {LLM} agents}.
\newblock In \emph{The Thirteenth International Conference on Learning Representations}.

\bibitem[{Tian and Goel(2025)}]{tian2025direct}
Thomas Tian and Kratarth Goel. 2025.
\newblock \href {https://openreview.net/forum?id=8UFG9D8xeU} {Direct multi-agent motion generation preference alignment with implicit feedback from demonstrations}.
\newblock In \emph{The Thirteenth International Conference on Learning Representations}.

\bibitem[{Wang et~al.(2024{\natexlab{a}})Wang, Zhang, Zhang, Wu, Mo, Lu, Wang, Li, Xu, Tang et~al.}]{wang2024comprehensive}
Fali Wang, Zhiwei Zhang, Xianren Zhang, Zongyu Wu, Tzuhao Mo, Qiuhao Lu, Wanjing Wang, Rui Li, Junjie Xu, Xianfeng Tang, et~al. 2024{\natexlab{a}}.
\newblock A comprehensive survey of small language models in the era of large language models: Techniques, enhancements, applications, collaboration with llms, and trustworthiness.
\newblock \emph{arXiv preprint arXiv:2411.03350}.

\bibitem[{Wang and Zhou(2024)}]{wang2024chain}
Xuezhi Wang and Denny Zhou. 2024.
\newblock Chain-of-thought reasoning without prompting.
\newblock \emph{arXiv preprint arXiv:2402.10200}.

\bibitem[{Wang et~al.(2022)Wang, Kordi, Mishra, Liu, Smith, Khashabi, and Hajishirzi}]{wang2022self}
Yizhong Wang, Yeganeh Kordi, Swaroop Mishra, Alisa Liu, Noah~A Smith, Daniel Khashabi, and Hannaneh Hajishirzi. 2022.
\newblock Self-instruct: Aligning language models with self-generated instructions.
\newblock \emph{arXiv preprint arXiv:2212.10560}.

\bibitem[{Wang et~al.(2024{\natexlab{b}})Wang, Mao, Fried, and Neubig}]{wang2024agent}
Zora~Zhiruo Wang, Jiayuan Mao, Daniel Fried, and Graham Neubig. 2024{\natexlab{b}}.
\newblock Agent workflow memory.
\newblock \emph{arXiv preprint arXiv:2409.07429}.

\bibitem[{Wu et~al.(2024)Wu, Wang, Yu, Zhang, Chang, and Yu}]{wu2024longmemeval}
Di~Wu, Hongwei Wang, Wenhao Yu, Yuwei Zhang, Kai-Wei Chang, and Dong Yu. 2024.
\newblock Longmemeval: Benchmarking chat assistants on long-term interactive memory.
\newblock \emph{arXiv preprint arXiv:2410.10813}.

\bibitem[{Xie et~al.(2024)Xie, Zhang, Chen, Li, Zhao, Cao, Hua, Cheng, Shin, Lei et~al.}]{xie2024osworld}
Tianbao Xie, Danyang Zhang, Jixuan Chen, Xiaochuan Li, Siheng Zhao, Ruisheng Cao, Toh~Jing Hua, Zhoujun Cheng, Dongchan Shin, Fangyu Lei, et~al. 2024.
\newblock Osworld: Benchmarking multimodal agents for open-ended tasks in real computer environments.
\newblock \emph{arXiv preprint arXiv:2404.07972}.

\bibitem[{Xie et~al.(2023)Xie, Zhou, Cheng, Shi, Weng, Liu, Hua, Zhao, Liu, Liu et~al.}]{xie2023openagents}
Tianbao Xie, Fan Zhou, Zhoujun Cheng, Peng Shi, Luoxuan Weng, Yitao Liu, Toh~Jing Hua, Junning Zhao, Qian Liu, Che Liu, et~al. 2023.
\newblock Openagents: An open platform for language agents in the wild.
\newblock \emph{arXiv preprint arXiv:2310.10634}.

\bibitem[{Xu et~al.(2024{\natexlab{a}})Xu, Kordi, Sanders, Wang, Byerly, Zhang, Durme, and Khashabi}]{DBLP:journals/corr/abs-2403-11905}
Kevin Xu, Yeganeh Kordi, Kate Sanders, Yizhong Wang, Adam Byerly, Jack Zhang, Benjamin~Van Durme, and Daniel Khashabi. 2024{\natexlab{a}}.
\newblock \href {https://doi.org/10.48550/arXiv.2403.11905} {Tur[k]ingbench: A challenge benchmark for web agents}.
\newblock \emph{CoRR}, abs/2403.11905.

\bibitem[{Xu et~al.(2024{\natexlab{b}})Xu, Han, Wang, Le, Madeka, Li, Wang, Agarwal, Lee, and Pfister}]{xu2024speculative}
Wenda Xu, Rujun Han, Zifeng Wang, Long~T Le, Dhruv Madeka, Lei Li, William~Yang Wang, Rishabh Agarwal, Chen-Yu Lee, and Tomas Pfister. 2024{\natexlab{b}}.
\newblock Speculative knowledge distillation: Bridging the teacher-student gap through interleaved sampling.
\newblock \emph{arXiv preprint arXiv:2410.11325}.

\bibitem[{Xu et~al.(2024{\natexlab{c}})Xu, Li, Tao, Shen, Cheng, Li, Xu, Tao, and Zhou}]{xu2024survey}
Xiaohan Xu, Ming Li, Chongyang Tao, Tao Shen, Reynold Cheng, Jinyang Li, Can Xu, Dacheng Tao, and Tianyi Zhou. 2024{\natexlab{c}}.
\newblock \href {https://arxiv.org/abs/2402.13116} {A survey on knowledge distillation of large language models}.
\newblock \emph{Preprint}, arXiv:2402.13116.

\bibitem[{Xu et~al.(2025)Xu, Lu, Shen, Wang, Wang, Mao, Xiong, and Yu}]{xu2025agenttrek}
Yiheng Xu, Dunjie Lu, Zhennan Shen, Junli Wang, Zekun Wang, Yuchen Mao, Caiming Xiong, and Tao Yu. 2025.
\newblock \href {https://openreview.net/forum?id=EEgYUccwsV} {Agenttrek: Agent trajectory synthesis via guiding replay with web tutorials}.
\newblock In \emph{The Thirteenth International Conference on Learning Representations}.

\bibitem[{Yang et~al.(2024{\natexlab{a}})Yang, Jimenez, Wettig, Lieret, Yao, Narasimhan, and Press}]{yang2024swe}
John Yang, Carlos~E Jimenez, Alexander Wettig, Kilian Lieret, Shunyu Yao, Karthik Narasimhan, and Ofir Press. 2024{\natexlab{a}}.
\newblock Swe-agent: Agent-computer interfaces enable automated software engineering.
\newblock \emph{arXiv preprint arXiv:2405.15793}.

\bibitem[{Yang et~al.(2025)Yang, Jimenez, Zhang, Lieret, Yang, Wu, Press, Muennighoff, Synnaeve, Narasimhan, Yang, Wang, and Press}]{yang2025swebench}
John Yang, Carlos~E Jimenez, Alex~L Zhang, Kilian Lieret, Joyce Yang, Xindi Wu, Ori Press, Niklas Muennighoff, Gabriel Synnaeve, Karthik~R Narasimhan, Diyi Yang, Sida Wang, and Ofir Press. 2025.
\newblock \href {https://openreview.net/forum?id=riTiq3i21b} {{SWE}-bench multimodal: Do {AI} systems generalize to visual software domains?}
\newblock In \emph{The Thirteenth International Conference on Learning Representations}.

\bibitem[{Yang et~al.(2024{\natexlab{b}})Yang, Liu, Chaudhary, Fakoor, Chaudhari, Karypis, and Rangwala}]{yang2024agentoccam}
Ke~Yang, Yao Liu, Sapana Chaudhary, Rasool Fakoor, Pratik Chaudhari, George Karypis, and Huzefa Rangwala. 2024{\natexlab{b}}.
\newblock \href {https://arxiv.org/abs/2410.13825} {Agentoccam: A simple yet strong baseline for llm-based web agents}.
\newblock \emph{Preprint}, arXiv:2410.13825.

\bibitem[{Yao et~al.(2023{\natexlab{a}})Yao, Chen, Yang, and Narasimhan}]{yao2023webshop}
Shunyu Yao, Howard Chen, John Yang, and Karthik Narasimhan. 2023{\natexlab{a}}.
\newblock \href {https://arxiv.org/abs/2207.01206} {Webshop: Towards scalable real-world web interaction with grounded language agents}.
\newblock \emph{Preprint}, arXiv:2207.01206.

\bibitem[{Yao et~al.(2023{\natexlab{b}})Yao, Zhao, Yu, Du, Shafran, Narasimhan, and Cao}]{yao2023react}
Shunyu Yao, Jeffrey Zhao, Dian Yu, Nan Du, Izhak Shafran, Karthik Narasimhan, and Yuan Cao. 2023{\natexlab{b}}.
\newblock React: synergizing reasoning and acting in language models (2022).
\newblock \emph{arXiv preprint arXiv:2210.03629}.

\bibitem[{Yin et~al.(2024)Yin, Brahman, Ravichander, Chandu, Chang, Choi, and Lin}]{yin2024agent}
Da~Yin, Faeze Brahman, Abhilasha Ravichander, Khyathi Chandu, Kai-Wei Chang, Yejin Choi, and Bill~Yuchen Lin. 2024.
\newblock Agent lumos: Unified and modular training for open-source language agents.
\newblock In \emph{Proceedings of the 62nd Annual Meeting of the Association for Computational Linguistics (Volume 1: Long Papers)}, pages 12380--12403.

\bibitem[{Yu and Lu(2024)}]{yu2024adam}
Shu Yu and Chaochao Lu. 2024.
\newblock Adam: An embodied causal agent in open-world environments.
\newblock \emph{arXiv preprint arXiv:2410.22194}.

\bibitem[{Yu et~al.(2025{\natexlab{a}})Yu, Peng, Vajipey, Cheng, Galley, Gao, and Yu}]{yu2025exact}
Xiao Yu, Baolin Peng, Vineeth Vajipey, Hao Cheng, Michel Galley, Jianfeng Gao, and Zhou Yu. 2025{\natexlab{a}}.
\newblock \href {https://arxiv.org/abs/2410.02052} {Exact: Teaching ai agents to explore with reflective-mcts and exploratory learning}.
\newblock \emph{Preprint}, arXiv:2410.02052.

\bibitem[{Yu et~al.(2025{\natexlab{b}})Yu, Peng, Vajipey, Cheng, Galley, Gao, and Yu}]{yu2025improving}
Xiao Yu, Baolin Peng, Vineeth Vajipey, Hao Cheng, Michel Galley, Jianfeng Gao, and Zhou Yu. 2025{\natexlab{b}}.
\newblock \href {https://openreview.net/forum?id=GBIUbwW9D8} {Improving autonomous {AI} agents with reflective tree search and self-learning}.
\newblock In \emph{The Thirteenth International Conference on Learning Representations}.

\bibitem[{Yuan et~al.(2025)Yuan, Bian, Li, Zhang, Guan, and Yu}]{yuan2025efficient}
Lei Yuan, Yuqi Bian, Lihe Li, Ziqian Zhang, Cong Guan, and Yang Yu. 2025.
\newblock \href {https://openreview.net/forum?id=EpnZEzYDUT} {Efficient multi-agent offline coordination via diffusion-based trajectory stitching}.
\newblock In \emph{The Thirteenth International Conference on Learning Representations}.

\bibitem[{Zeng et~al.(2023)Zeng, Liu, Lu, Wang, Liu, Dong, and Tang}]{zeng2023agenttuning}
Aohan Zeng, Mingdao Liu, Rui Lu, Bowen Wang, Xiao Liu, Yuxiao Dong, and Jie Tang. 2023.
\newblock \href {https://arxiv.org/abs/2310.12823} {Agenttuning: Enabling generalized agent abilities for llms}.
\newblock \emph{Preprint}, arXiv:2310.12823.

\bibitem[{Zhang et~al.(2024)Zhang, Ma, Ma, Han, Wu, and Tresp}]{zhang2024webpilot}
Yao Zhang, Zijian Ma, Yunpu Ma, Zhen Han, Yu~Wu, and Volker Tresp. 2024.
\newblock \href {https://arxiv.org/abs/2408.15978} {Webpilot: A versatile and autonomous multi-agent system for web task execution with strategic exploration}.
\newblock \emph{Preprint}, arXiv:2408.15978.

\bibitem[{Zhang and Zhang(2023)}]{zhang2023you}
Zhuosheng Zhang and Aston Zhang. 2023.
\newblock You only look at screens: Multimodal chain-of-action agents.
\newblock \emph{arXiv preprint arXiv:2309.11436}.

\bibitem[{Zhou et~al.(2023{\natexlab{a}})Zhou, Yan, Shlapentokh-Rothman, Wang, and Wang}]{zhou2023language}
Andy Zhou, Kai Yan, Michal Shlapentokh-Rothman, Haohan Wang, and Yu-Xiong Wang. 2023{\natexlab{a}}.
\newblock Language agent tree search unifies reasoning acting and planning in language models.
\newblock \emph{arXiv preprint arXiv:2310.04406}.

\bibitem[{Zhou et~al.(2023{\natexlab{b}})Zhou, Xu, Zhu, Zhou, Lo, Sridhar, Cheng, Ou, Bisk, Fried et~al.}]{zhou2023webarena}
Shuyan Zhou, Frank~F Xu, Hao Zhu, Xuhui Zhou, Robert Lo, Abishek Sridhar, Xianyi Cheng, Tianyue Ou, Yonatan Bisk, Daniel Fried, et~al. 2023{\natexlab{b}}.
\newblock Webarena: A realistic web environment for building autonomous agents.
\newblock \emph{arXiv preprint arXiv:2307.13854}.

\end{thebibliography}
